\documentclass[twocolumn]{aastex631}

\newcommand{\todo}[1]{\textcolor{black}{#1}}

\newcommand{\angstrom}{\r{A}} 

\newcommand\HII{$\textrm{H}\scriptstyle\mathrm{II}$}

\shorttitle{Slow strong bars affect their hosts the most}
\shortauthors{G\'eron et al.}
\graphicspath{{./}{figures/}}

\begin{document}

\title{The effects of bar strength and kinematics on galaxy evolution: \\slow strong bars affect their hosts the most}

\correspondingauthor{Tobias G\'eron}
\email{tobias.geron@utoronto.ca}

\author[0000-0002-6851-9613]{Tobias G\'eron}
\affiliation{Dunlap Institute for Astronomy \& Astrophysics, University of Toronto \\
50 St. George Street \\
Toronto, ON M5S 3H4, Canada}
\affiliation{Oxford Astrophysics, Department of Physics, University of Oxford \\
Denys Wilkinson Building, Keble Road \\
Oxford, OX1 3RH, UK}

\author[0000-0001-6417-7196]{R. J. Smethurst}
\affiliation{Oxford Astrophysics, Department of Physics, University of Oxford \\
Denys Wilkinson Building, Keble Road \\
Oxford, OX1 3RH, UK}

\author[0000-0001-5578-359X]{Chris Lintott}
\affiliation{Oxford Astrophysics, Department of Physics, University of Oxford \\
Denys Wilkinson Building, Keble Road \\
Oxford, OX1 3RH, UK}

\author[0000-0003-0846-9578]{Karen L. Masters}
\affiliation{Haverford College, Department of Physics and Astronomy \\
370 Lancaster Avenue \\
Haverford, Pennsylvania 19041, USA}

\author[0000-0002-3887-6433]{I. L. Garland}
\affiliation{Department of Physics, Lancaster University \\
Lancaster, LA1 4YB, UK
}

\author{Petra Mengistu}
\affiliation{Haverford College, Department of Physics and Astronomy \\
370 Lancaster Avenue \\
Haverford, Pennsylvania 19041, USA}

\author[0000-0003-1217-4617]{David O'Ryan}
\affiliation{Department of Physics, Lancaster University \\
Lancaster, LA1 4YB, UK
}

\author[0000-0001-5882-3323]{B.D. Simmons}
\affiliation{Department of Physics, Lancaster University \\
Lancaster, LA1 4YB, UK
}



\begin{abstract}

  We study how bar strength and bar kinematics affect star formation in different regions of the bar by creating radial profiles of EW[H$\alpha$] and D$_{\rm n}$4000 using data from SDSS-IV MaNGA. Bars in galaxies are classified as strong or weak using Galaxy Zoo DESI, and they are classified as fast and slow bars using the Tremaine-Weinberg method on stellar kinematic data from the MaNGA survey. In agreement with previous studies, we find that strong bars in star forming galaxies have enhanced star formation in their centre and beyond the bar-end region, while star formation is suppressed in the arms of the bar. This is not found for weakly barred galaxies, which have very similar radial profiles to unbarred galaxies. In addition, we find that slow bars in star forming galaxies have significantly higher star formation along the bar than fast bars. However, the global star formation rate is not significantly different between galaxies with fast and slow bars. This suggests that the kinematics of the bar do not affect star formation globally, but changes where star formation occurs in the galaxy. Thus, we find that a bar will influence its host the most if it is both strong and slow.

\end{abstract}

\keywords{Galaxy bars (2364) --- Galaxy evolution (594) --- Galaxy kinematics (602) --- Star formation (1569)}


\section{Introduction}

Bars are common structures among disc galaxies, as 44-52\% of low-redshift disc galaxies observed with optical wavelengths have a clearly defined bar \citep{marinova_2007,barazza_2008,aguerri_2009,buta_2019}. This fraction goes up to 59-73\% when using infrared wavelengths \citep{eskridge_2000,marinova_2007,menendez_delmestre_2007,sheth_2008}, presumably due to infrared wavelengths not being as affected by dust and star formation \citep{erwin_2018}. Other studies report lower bar fractions, ranging between 24-29\% (e.g. see \citealp{masters_2011, skibba_2012, cheung_2013}), although these studies noted that they primarily consider strong bars and leave out weak bars. The bar fraction is also dependent on the stellar mass of the galaxy. For example, \citet{lazar_2024} have found a bar fraction of 11\% for dwarf galaxies. Finally, the bar fraction among disc galaxies at higher redshifts (0.5 $<$ $z$ $<$ 2) is also lower, ranging between 10-23\% \citep{elmegreen_2004,sheth_2008,melvin_2014,simmons_2014}. Bars are known to influence their host in multiple ways. As a bar grows over time, it will transfer angular momentum outwards from the inner disc to the outer disc and the dark matter halo \citep{lynden_bell_1972,sellwood_1981,athanassoula_2003,athanassoula_2013}. The growing bar will also funnel gas to the centre of the galaxy along the arms of the bar \citep{sorensen_1976,athanassoula_1992,davoust_2004,villa-vargas_2010,vera_2016,spinoso_2017,george_2019}. Multiple studies have found that bars appear more often in massive, red and gas-poor galaxies (i.e. typical quiescent galaxies, \citealp{hoyle_2011,masters_2011, masters_2012,cheung_2013,cervantessodi_2017,vera_2016,kruk_2018,fraser_mckelvie_2020b}), which suggests that bars are linked to the quenching process. One proposed quenching mechanism states that bars can trigger a starburst in the centre of the galaxy after causing substantial inflow of gas, which increases the rate of gas consumption and eventually quenches the host \citep{alonso_herrero_2001, sheth_2005, jogee_2005, hunt_2008, carles_2016}. Another possibility is that bars increase the velocity dispersion or shear in the bar region, so that the gas becomes too dynamically hot for star formation \citep{athanassoula_1992,reynaud_1998,sheth_2000,zurita_2004, haywood_2016,khoperskov_2018,hogarth_2024}. 

There is a large variety in bar length and strength, which is typically addressed by classifying bars as either weak or strong. This terminology goes back to \citet{devaucouleurs_1959, devaucouleurs_1963}, who defined strong bars as obvious and long structures, whereas weak bars are faint and small structures. There are many ways to measure bar strength. For example, the maximum ellipticity \citep{athanassoula_1992b,laurikainen_2002,erwin_2004} and the boxiness of the bar isophotes \citep{gadotti_2011} have been used to estimate bar strength, as galaxies with stronger bars have more pronounced and elongated isophotes. One can also approximate the bar strength by measuring the torque exerted by the bar \citep{combes_1981, buta_2001, laurikainen_2002, speltincx_2008} or by estimating the amplitude of the m=2 Fourier mode \citep{athanassoula_2003,garcia_gomez_2017}. In the catalogue of detailed visual morphological classifications of \citet{nair_2010}, a bar is strong if it dominates the light distribution of the galaxy and vice versa for a weak bar. The Galaxy Zoo 2 project \citep{willett_2013} explored another method to measure bar strength. They asked citizen scientists to identify bars on images of the Sloan Digital Sky Survey (SDSS, \citealp{blanton_2017}) and noted that bar strength is correlated with the amount of volunteers that voted that there is a bar in the galaxy \citep{masters_2011, skibba_2012}. Galaxy Zoo DECaLS \citep{walmsley_2022} expanded on this idea and asked citizen scientists to classify bars into strong or weak based on images from Dark Energy Camera Legacy Survey (DECaLS, \citealp{dey_2019}). These classifications were used in \citet{geron_2021}, who found that differences between weak and strong bars disappeared when correcting for bar length. They suggested that weak and strong bars are not fundamentally different physical phenomena, but rather that weak and strong bars are part of a continuum of bar types, which changes from `weakest' to `strongest'.

While morphology is often used as a proxy for kinematics, it is also important to look at the kinematics of galaxies, rather than morphology alone, as insight can be derived from considering both. This has been demonstrated by \citet{emsellem_2011} and \citet{cappellari_2011}, who found that early-type galaxies with similar morphologies can be classified into two distinct types, fast and slow rotators, based on their kinematics. Slow rotators are dispersion dominated galaxies that have undergone many gas-poor mergers. In contrast, fast rotators are rotationally supported structures that slowly build up their bulge by accreting gas \citep{emsellem_2011,bois_2011,duc_2011,naab_2014,cappellari_2016}. The kinematics of bars has also been shown to be important. The bar pattern speed ($\Omega_{\rm bar}$), also known as the rotational frequency of the bar, is correlated with the evolution of the bar and its host in simulations. As the bar grows and exchanges angular momentum with its host, the bar typically slows down and the pattern speed decreases \citep{debattista_2000,athanassoula_2003, martinez_valpuesta_2006,okamoto_2015}. Another important parameter is the corotation radius (R$_{\rm CR}$), which is the radius at which the stars have the same angular speed as the pattern speed of the bar. The corotation radius is used to calculate the dimensionless ratio $\mathcal{R}$, which is defined as $\mathcal{R} = R_{\rm CR} / R_{\rm bar}$, where $R_{\rm bar}$ is the deprojected bar radius. This ratio is used to classify bars into slow ($\mathcal{R} > 1.4$), fast ($1.0 < \mathcal{R} < 1.4$) and ultrafast ($\mathcal{R} < 1.0$) bars (e.g. see \citealp{debattista_2000,rautiainen_2008,aguerri_2015}). In simulations, $\mathcal{R}$ is thought to be correlated with the formation mechanism of the bar. Bars triggered by tidal interactions tend to remain slow for a longer time and have overall higher values of $\mathcal{R}$ compared to bars formed by global bar instabilities \citep{sellwood_1981,miwa_1998,martinez_valpuesta_2016,martinez_valpuesta_2017}.

\begin{figure}
\includegraphics[width=\columnwidth]{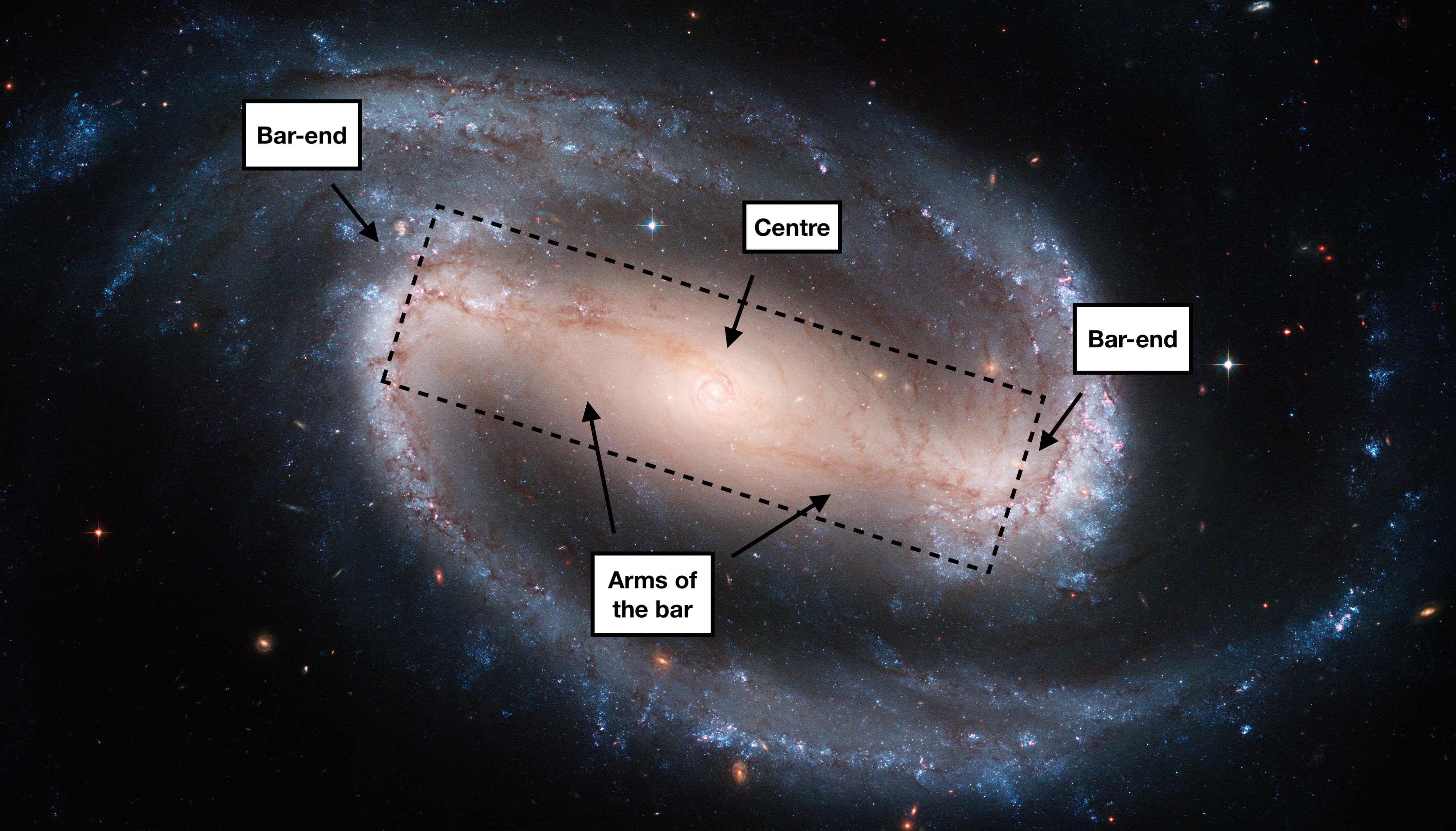}
\caption{NGC 1300 is arguably the most famous barred galaxy. Different important regions of a bar are highlighted in this figure. The centre of the bar usually coincides with the centre of the galaxy. The arms of the bar connect the centre with the bar-end regions. The dust lanes along the arms of the bar are clearly visible and the spiral arms connect directly the the bar-ends. The dashed rectangle delineates the barred region. Credits: NASA, ESA, and The Hubble Heritage Team (STScI/AURA).} 
\label{fig:ngc1300}
\end{figure}

In this paper, we will consistently refer to distinct regions within barred galaxies, as illustrated in Figure \ref{fig:ngc1300} using the well-known barred galaxy NGC 1300. This figure shows a very prominent bar that connects to the two spiral arms of the galaxy. The key regions in the bar are the centre, the arms of the bar and the bar-end region. The region between the two bar-ends is referred to as the `barred region', while everything else is considered `outside the bar'. While the centre of the bar often coincides with the centre of the galaxy, it is worth noting that offset bars have been observed in a limited number of galaxies. According to \citet{kruk_2017}, only 2\% of galaxies with masses comparable to the Milky Way ($10^{10.5} - 10^{11.1}$ M$_{\odot}$) have offset bars.

We know from observations that barred galaxies have more star formation in their centre \citep{alonso_herrero_2001, hunt_2008, ellison_2011,coelho_2011,lin_2020} and bar-end regions \citep{reynaud_1998,verley_2007,diaz_garcia_2020,fraser_mckelvie_2020,maeda_2020b}, while they suppress star formation along the arms of the bar \citep{reynaud_1998,zurita_2004,watanabe_2011,haywood_2016}. These observations show that bars play a role in the evolution of their host galaxies, although it is unclear whether this is true for both weak and strong bars, and for both fast and slow bars. \citet{geron_2021} looked into the effect of strong and weak bars and showed that strong bars have higher central SFRs, while weak bars do not. However, the influence of bar strength on star formation in the arms of the bar and the bar-end region has not yet been studied in detail. It is also uncertain how the kinematics of bars affect galaxy evolution and quenching. \citet{geron_2023} classified a sample of bars as either fast ($1.0 < \mathcal{R} < 1.4$) or slow ($\mathcal{R} > 1.4$), which we use in this paper to address the following questions. Do fast and slow bars affect star formation in the same way? Do the kinematics of a bar contribute to how strong or weak bars affect their host galaxies? Or in other words, do slow strong bars and fast strong bars affect their hosts in the same way? To answer these questions, we make use of resolved star formation indicators obtained through the Mapping Nearby Galaxies at Apache Point Observatory (MaNGA) survey \citep{bundy_2015}, which allows us to study different regions within barred galaxies. In Section \ref{sec:r_profile_sb_wb}, we construct radial profiles of EW[H$\alpha$] and D$_{\rm n}$4000 to assess differences in star formation between strong and weak bars. A comparable analysis is carried out for fast and slow bars in Section \ref{sec:r_profile_fb_sb}. The results are discussed in Section \ref{sec:discussion} and the main conclusions are summarised in Section \ref{sec:r_profile_conclusions}. We assume a standard flat $\Lambda$CDM cosmological model with H$_{0}$ = 70 km s$^{-1}$ Mpc$^{-1}$, $\Omega_{\rm m}$ = 0.3 and $\Omega_{\rm \Lambda}$ = 0.7 where necessary.

\section{Data and Methods}

\subsection{Galaxy Zoo}
\label{sec:galaxy_zoo}

We used the Galaxy Zoo (GZ) project to identify weak and strong bars. In GZ, citizen scientists are shown images of galaxies and answer multiple questions about their morphology according to a decision tree structure \citep{lintott_2008, lintott_2011}. This paper makes use of Galaxy Zoo DESI (GZ DESI, \citealp{walmsley_2023}), which used images obtained from the DESI Legacy Imaging Surveys\footnote{\url{www.legacysurvey.org/}} \citep{dey_2019}. DESI consists of three individual projects: the Dark Energy Camera Legacy Survey (DECaLS), the Beijing-Arizona Sky Survey (BASS) and the Mayall z-band Legacy Survey (MzLS) that collectively cover $\sim14,000$ deg$^{2}$ of sky. The DESI Legacy Imaging Surveys are relatively deep (e.g. the PSF depth of the $r$-band is 23.54 for DECaLS, \citealp{dey_2019}), which means that it is possible to reliably identify weak bars in these images. The classifications of citizen scientists are used in GZ DESI to train machine classifications based on Bayesian convolutional neural networks described in \citet{walmsley_2022}. We use these machine classifications in this paper to classify bars as strong or weak (more details are provided in Section \ref{sec:sample_selection}).

\subsection{MaNGA}
\label{sec:manga}

To investigate how bars impact their host galaxies, we make use of H$\alpha$ and D$_{\rm n}$4000 measurements obtained from the MaNGA survey \citep{bundy_2015}, which is part of the Sloan Digital Sky Survey IV (SDSS-IV) collaboration \citep{blanton_2017}. In this paper, we used data from the seventeenth data release of SDSS \citep{abdurrouf_2022}. MaNGA used the Baryon Oscillation Spectroscopic Survey (BOSS) spectrograph, which has a wavelength coverage of 3600 - 10,000 \AA\ and a resolution of R $\sim 2000$ \citep{smee_2013}, on the 2.5m Sloan Telescope at Apache Point Observatory \citep{gunn_2006}. MaNGA is an integral field unit (IFU) survey, which means that it stacks 19-127 optical fibers hexagonally to obtain spectra at multiple positions for a galaxy. Galaxies are typically covered out to 1.5 effective radii (R$_{e}$), although a third are covered up until to 2.5 R$_{e}$ \citep{wake_2017}. In this paper, we use the Voronoi binned maps that are binned to S/N $\sim$10 \citep{westfall_2019}. More information on the observing strategy, survey design, data reduction process, sample selection and the data analysis pipeline can be found in \citet{law_2015,yan_2016,law_2016,wake_2017, belfiore_2019, westfall_2019}. The SFRs and stellar masses used in this paper are obtained from the Pipe3D value added catalog \citep{sanchez_2016,sanchez_2016b}. The SFRs in Pipe3D are based on the H$\alpha$ flux and are corrected for dust extinction. More details can be found in \citet{sanchez_2016b}. Additionally, we made extensive use of the \textit{Marvin} software to access the MaNGA data \citep{cherinka_2019}.

We use the Gaussian-fitted equivalent width measurement of H$\alpha$, EW[H$\alpha$], in order to investigate the impact of bars on their host galaxies. H$\alpha$ emission, found at a rest-frame wavelength of 6564~\angstrom{}, originates from \HII{} regions that are excited by the ionizing radiation emitted by OB type stars found in young stellar populations \citep{argence_2009, spindler_2018, smethurst_2019}. The amount of H$\alpha$ radiation is proportional to the number of OB stars and, consequently, the amount of star formation \citep{kennicutt_1998}. We also look at the strength of the 4000~\angstrom{} break, which is caused by the combined effects of absorption around 4000~\angstrom{} by the metals in the atmosphere of older and cooler stars and the absence of emission from young, blue OB type stars. This implies that the 4000~\angstrom{} break traces the age of the stellar population, although there is a dependence on the stellar metallicity \citep{kauffmann_2003, smethurst_2019, paulino_afonso_2020}. We use the definition of the 4000~\angstrom{} break from \citet{balogh_1999}, known as D$_{\rm n}$4000, which is an update of the definition proposed by \citet{bruzual_1983}. D$_{\rm n}$4000 is a dimensionless ratio of the flux between 4000-4100~\angstrom{} and 3850-3950~\angstrom{}, as measured by the MaNGA data analysis pipeline (DAP, \citealp{westfall_2019}). A low D$_{\rm n}$4000 value suggests recent star formation, indicating a relatively young average stellar population. Conversely, an older stellar population is present if D$_{\rm n}$4000 is high \citep{kauffmann_2003, smethurst_2019, paulino_afonso_2020}. This also means that D$_{\rm n}$4000 serves as a useful probe to assess whether a quenching mechanism (e.g. a strong bar) has had a long-lasting effect on the galaxy, which helps to probe the timescales of that mechanism.

\subsection{Tremaine-Weinberg method and bar kinematics}
\label{sec:tw_method}

The Tremaine-Weinberg (TW) method is a model-independent method to calculate the pattern speed of a galaxy \citep{tremaine_1984}. Its main assumptions are that the tracer used to calculate the pattern speed (i.e. stars or gas) satisfies the continuity equation and that there is a well-defined pattern speed. The TW method has been used frequently to measure the bar pattern speed (e.g. see \citealp{aguerri_2015,cuomo_2019,guo_2019,garma_oehmichen_2020,garma_oehmichen_2022}). More recently, \citet{geron_2023} used the TW method to obtain reliable measurements of the bar pattern speed, corotation radius and $\mathcal{R}$ for a sample of \todo{210} barred galaxies that have a wide range of bar strengths. The TW method can be written as: 
\begin{equation}
    \Omega_{\rm b} \sin{\left( i \right)} = \frac{\left<V\right>}{\left<X\right>} \;,
\end{equation}
where $\Omega_{\rm b}$ is the bar pattern speed, $i$ is the inclination of the galaxy, $\left<X\right>$ is the photometric integral and $\left<V\right>$ is the kinematic integral. These last two integrals are defined as: 

\begin{equation}
  \left<X\right> = \frac{\int_{\rm -\infty}^{\rm +\infty} X \Sigma (X,Y) \text{d}\Sigma}{\int_{\rm -\infty}^{\rm +\infty} \Sigma (X,Y) \text{d}\Sigma} \;,
  \label{eq:tw_X_eqs}
\end{equation}

\begin{equation}
  \left<V\right> = \frac{\int_{\rm -\infty}^{\rm +\infty} V_{\rm LOS} (X,Y) \Sigma (X,Y) \text{d}\Sigma}{\int_{\rm -\infty}^{\rm +\infty} \Sigma (X,Y) \text{d}\Sigma} \; ,
  \label{eq:tw_V_eqs}
\end{equation}
where $\Sigma$ is the surface brightness of the galaxy and $V_{\rm LOS}$ is the line of sight velocity of the galaxy. The coordinate system $\left(X,Y\right)$ is found in the sky plane with the origin centered on the centre of the galaxy and the $X$-axis aligned with the major axis of the galaxy. \citet{geron_2023} used surface brightness and line of sight velocity data from the MaNGA survey. The integration is done along multiple pseudo-slits placed parallel to the major axis of the galaxy across the IFU to make sure that the final measurement of the pattern speed is reliable. For more details on the technical aspects of the TW method, refer to Section 2.1 in \citet{geron_2023}. 

Once the pattern speed is calculated, it is possible to find the corotation radius ($R_{\rm CR}$) if the rotation curve of the galaxy is known. The corotation radius is the radius at which the gravitational and centrifugal forces balance each other in the rest frame of the bar. This means that, at the corotation radius, the stars in the disc will have the same angular velocity as the bar. In \citet{geron_2023}, the rotation curve of the galaxy is determined by fitting the stellar velocity data of MaNGA to a two parameter arctan function, described in \citet{courteau_1997}:
\begin{equation}
    V_{\rm rot} = V_{\rm sys} + \frac{2}{\pi} V_{\rm c} \arctan{\left(\frac{r - r_{0}}{r_{t}}\right)}\;,
\end{equation}
where V$_{\rm rot}$ is the true stellar velocity (i.e. corrected for the inclination of the galaxy), V$_{\rm sys}$ is the systemic velocity of the galaxy, V$_{\rm c}$ is the asymptotic velocity, $r$ is the deprojected distance to the centre of the galaxy, r$_{0}$ is the spatial centre of the galaxy and r$_{t}$ is the transition radius. In this model, the rotation curve flattens at r$_{t}$ and goes towards V$_{\rm c}$. As the corotation radius is defined as the radius where the stars in the disc rotate with the same velocity as the bar, it can be identified by finding the intersection of the rotation curve and the line found by multiplying the bar pattern speed by the distance from the centre of the galaxy (i.e. $\Omega_{\rm b} \cdot $ r, see Figure \todo{4} of \citealt{geron_2023}). 

Finally, the dimensionless parameter $\mathcal{R}$ can be calculated with $\mathcal{R} = R_{\rm CR} / R_{\rm bar}$, where $R_{\rm bar}$ is the deprojected bar radius. This parameter is used to classify bars into slow ($\mathcal{R} > 1.4$) and fast bars ($1.0 < \mathcal{R} < 1.4$) \citep{debattista_2000,rautiainen_2008,aguerri_2015}. Although bars with $\mathcal{R}$ $<$ 1.0, which are called ultrafast bars, should theoretically not exist, as bar orbits become unstable beyond the corotation radius \citep{contopoulos_1980,contopoulos_1981, athanassoula_1992}, they have been nevertheless repeatedly observed \citep{buta_2009, aguerri_2015, cuomo_2019,guo_2019,garma_oehmichen_2020}. Slow bars have bar lengths that are shorter than the corotation radius, whereas fast bars end near the corotation radius. This also implies that the bar-ends of a fast bar will rotate with a velocity similar to the stars in the disc at that radius. In contrast, the bar-ends of slow bars rotate much slower than the stars in the neighbouring parts of the disc. Refer to \citet{geron_2023} for more details and a step-by-step example on how $\Omega_{\rm b}$, the rotation curve, the corotation radius and $\mathcal{R}$ are calculated using the TW method. The code used to calculate these kinematic parameters and the full table of final values presented in \citet{geron_2023} are publicly available \href{https://doi.org/10.5281/zenodo.7567945}{here}\footnote{\url{https://doi.org/10.5281/zenodo.7567945}}. The fast and slow bars identified there are used throughout this paper.

\subsection{Radial profiles}
\label{sec:radius_profile_method}

\begin{figure*}
    \centering
     \includegraphics[width=\textwidth]{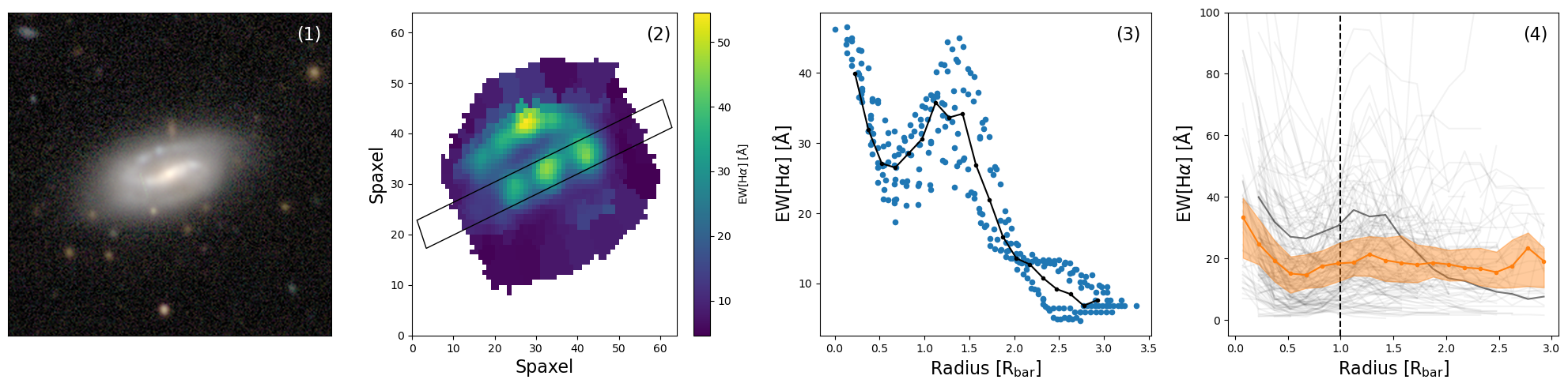}
     \caption{This figure demonstrates how the radial profiles are constructed using a randomly picked strongly barred star forming galaxy (plate-ifu: 12622-9101), the DECaLS images of which is shown in the first panel. The second panel shows the EW[H$\alpha$] map obtained from MaNGA. This panel is overlaid with an aperture positioned along the PA of the bar. All the spaxels whose centre lie within the aperture are extracted from the map. The EW[H$\alpha$] of those spaxels is plotted against the distance to the centre of the galaxy normalised to the bar radius in the third panel. The radial profile of this particular galaxy is shown by the black line. Finally, if multiple galaxies are binned in the same way, it is possible to create an median profile of any sample of galaxies by taking the median value in each bin. The fourth panel shows the radial profiles of all strongly barred star forming galaxies in the GZ DESI-MaNGA sample using the faint grey lines. The black line corresponds to the radial profile calculated in the third panel. The thicker orange line shows the median radial profile of EW[H$\alpha$] for all strongly barred star forming galaxies (n = \todo{172}). The shaded regions represent the 33$^{\rm rd}$ and 66$^{\rm th}$ percentile in every bin. The dashed line at $R_{\textrm{bar}}$ = 1 denotes the end of the bar.}
     \label{fig:r_profiles_intro}
 \end{figure*}

We will study how a bar influences different regions (such as the centre, the arms of the bar, and the bar-end) of a galaxy. This analysis involves creating radial profiles for each target over any particular map from MaNGA along any given position angle (e.g. the PA of the bar). A median profile is then created by taking the median of individual profiles belonging to a specific group (e.g. for all strongly barred galaxies). This process is shown in Figure \ref{fig:r_profiles_intro}. The leftmost panel shows a DECaLS image of a random strongly barred star forming (SF) galaxy. The second panel shows the Gaussian-fitted equivalent width measurement of H$\alpha$ (EW[H$\alpha$]) from MaNGA, with an overlay of an aperture with a width of \todo{3} arcsec. The aperture is usually positioned along the PA of the bar. However, in some cases, the aperture can be placed perpendicular to the bar in order to probe off-bar regions. The projected distance to the centre of the galaxy is normalised to the projected bar radius for each spaxel whose centre falls within the aperture. All the bar lengths and bar PAs used in this work were previously measured in \citet{geron_2023}; refer to their Section 3.3.3 for more details on these measurements. These normalised distances are then plotted against EW[H$\alpha$] in the third panel. Next, the radial profile for this galaxy is calculated by binning these spaxels into bins with a width of \todo{0.15} $R_{\textrm{bar}}$ and computing the median within each bin. This bin width was chosen as a compromise between generating detailed profiles and the sample size. However, the results of this paper do not change qualitatively if other reasonable bin widths (e.g. anywhere between \todo{0.10} $R_{\textrm{bar}}$ - \todo{0.3} $R_{\textrm{bar}}$) are used instead. The radial profile of this specific galaxy has an interesting shape. The EW[H$\alpha$] appears to peak at the centre, after which it decreases until $\sim$\todo{0.5} $R_{\textrm{bar}}$. Subsequently, the EW[H$\alpha$] rises again, reaching a second peak around $\sim$\todo{1.2} $R_{\textrm{bar}}$, after which it falls again. Interestingly, this second peak reaches its maximum right beyond the end of the bar. In this particular example, it seems that the EW[H$\alpha$] is high in the centre and bar-ends, while lower in the arms of the bar and outside the bar. It is also possible to construct radial profiles in terms of kpc, instead of normalising to the bar radius. The bin width in this case is \todo{0.5} kpc.

The faint grey lines in the background of the rightmost panel of Figure \ref{fig:r_profiles_intro} represent the median radial profiles for all strongly barred star forming galaxies within the GZ DESI-MaNGA sample (more details of this sample will be given in Section \ref{sec:sample_selection}). When galaxies are consistently binned in the same way, it is possible to generate a median profile for any given galaxy sample by calculating the median of all individual profiles in every bin. This median radial profile of EW[H$\alpha$], which is representative of the strongly barred star forming galaxies population in the GZ DESI-MaNGA sample, is shown by the thicker orange line.

In certain situations (e.g. to facilitate comparison with barred galaxies), we want to create radial profiles normalised to $R_{\textrm{bar}}$ for unbarred galaxies. To achieve this, we assign bar lengths to unbarred galaxies based on the their stellar masses. We first create stellar mass bins with a width of  0.25 $\log{ \left(\textrm{M}_{\textrm{*}}/\textrm{M}_{\odot} \right)}$. We then modelled the bar length distribution of the barred galaxies in our sample in each bin using a log-normal distribution. Each unbarred galaxy was then assigned a bar length in kpc to normalise with, which was randomly drawn from the distribution corresponding to its mass bin. However, it is crucial to note that this process was done just to  establish a reference point for comparing with barred galaxies and that any observed trends with $R_{\textrm{bar}}$ for unbarred galaxies do not have any physical meaning.

\subsection{Sample selection}
\label{sec:sample_selection}

There are two main samples used in this work. The first one is called the GZ DESI-MaNGA sample. This sample consists of the machine classifications of GZ DESI \citep{walmsley_2023}, which are cross-matched against MaNGA \citep{bundy_2015}. This cross-match contains a total of \todo{9,812} galaxies. We further apply thresholds on the redshift (0.01 $<$ $z$ $<$ 0.05, spectroscopic redshift estimates are from the NASA-Sloan Atlas) and the absolute $r$-band magnitude ($M_{r}$ $<$ -18.96, also from the NASA-Sloan Atlas), in order to remove targets where a potential bar is difficult to detect because it is too faint or too distant. This reduced the sample size to \todo{5,810} galaxies.

Additionally, volunteers in GZ DESI are only asked ``\textit{Is there a bar feature through the centre of the galaxy?}'' when they first answered that the galaxy is a disc galaxy that is not viewed edge-on. Additional thresholds should therefore be used to make sure enough volunteers answered the bar question and to guarantee reliable bar classifications \citep{willett_2013}. Since the machine classifications used here are trained on the volunteer classifications, similar thresholds should be considered. We place thresholds on the estimated fraction of volunteers that would have voted that the galaxy has features or a disc ($p_{\textrm{features/disk}} \geq 0.27$), the estimated fraction of volunteers that would have voted that the galaxy is not viewed edge-on ($p_{\textrm{not edge-on}} \geq 0.68$) and the estimated fraction of volunteers that would have been asked the bar question ($N_{\textrm{bar}}\geq 0.5$). More information on these thresholds can be found in \citet{geron_2021} and \citet{walmsley_2022}. These thresholds further reduced the sample size to \todo{2,125} galaxies.

Using the GZ DESI machine classifications, every galaxy was assigned one of three bar types: strong, weak or no bar. A galaxy is classified as unbarred if more than half of all the predicted classifications voted that the galaxy did not have a strong or weak bar, i.e. $p_{\textrm{strong bar + weak bar}} < 0.5$. If this was not the case, then the galaxy had a weak bar if $p_{\textrm{strong bar}} < p_{\textrm{weak bar}}$ and a strong bar if $p_{\textrm{strong bar}} \geq p_{\textrm{weak bar}}$. These thresholds have been successfully used before in \citet{geron_2021} and \citet{geron_2023}. This classification scheme results in a strong bar fraction of \todo{17.0}\% (\todo{363/2,125}) and a weak bar fraction of \todo{37.5}\% (796/2,125).

Finally, the galaxies are also divided into star forming (SF) and quenching using the star formation sequence defined by \citet{belfiore_2018}:

\begin{eqnarray}
  \label{eq:sfr_mass_split}
  \log{ \left( \textrm{SFR}/ \textrm{M}_{\odot}\,\textrm{yr}^{-1} \right) } = (0.73\,\pm\,0.03) \log{ \left(\textrm{M}_{\textrm{*}}/\textrm{M}_{\odot} \right)} \nonumber \\ - (7.33\,\pm\,0.29)\;.
\end{eqnarray}

We assume that any galaxy that is 1$\sigma$ (=0.39 dex) below this line are quenching and all other galaxies are SF. The GZ DESI-MaNGA sample is used to assess the effects of bar strength on different regions in the galaxy (see Sections \ref{sec:r_profile_sb_wb} and \ref{sec:disc_sbwb}).

The second sample used in this work is called the TW sample, which is a subsample of the GZ DESI-MaNGA sample. \citet{geron_2023} calculated the bar pattern speed, corotation radius and $\mathcal{R}$ for the galaxies in the GZ DESI-MaNGA sample using the TW method. However, there are a few limitations and additional thresholds that need to be put in place in order to use the TW method that limit the final sample size. To begin with, all unbarred galaxies are removed, which reduced the sample size to \todo{1,159} galaxies. The TW method can only be applied to galaxies with regular kinematics, which removed another \todo{474} galaxies. Additionally, the TW method only works for galaxies where the bar is not aligned with the disc major or minor axis and for galaxies with intermediate inclinations. These two restrictions removed another \todo{193} and \todo{32} galaxies, respectively. Additional thresholds on the quality of the TW output are used to ensure reliable estimates of the kinematic parameters. All these thresholds and restrictions are explained in much greater detail in \citet{geron_2023}. These thresholds resulted in a final sample of \todo{210} galaxies that have reliable measurements of the bar pattern speed, corotation radius and the ratio $\mathcal{R}$. 

Using this ratio, we find that \todo{67}\% of the bars in the TW sample are slow bars ($\mathcal{R} > 1.4$), \todo{21}\% are fast bars ($1.0 < \mathcal{R} < 1.4$) and \todo{12}\% are ultrafast bars ($\mathcal{R} < 1.0$). However, only \todo{2}\% of the ultrafast bars that \citet{geron_2023} identified in the TW sample are confidently within the ultrafast regime. Because only very few ultrafast bars exist in this sample, they are grouped together with the fast bars for the remainder of this work. The TW sample is used to study the effects of bar kinematics on different regions in the galaxy (see Sections \ref{sec:r_profile_fb_sb} and \ref{sec:disc_fbsb}).

\subsection{Contamination by AGN}
\label{sec:agn}

H$\alpha$ emission from star formation can be contaminated by other sources, such as active galactic nuclei (AGN), which can bias the results presented in this work. This is particularly important when studying bars, as some studies suggest that bar-driven inflow of gas could trigger AGN activity \citep{knapen_2000,oh_2012,fanali_2015,galloway_2015}. To ensure the results presented here are not biased by the presence of AGN, we cross-matched the galaxies in the GZ DESI-MaNGA catalogue to the MaNGA AGN value added catalogue \citep{comerford_2020}. The MaNGA AGN catalogue identified AGN in MaNGA galaxies using four distinct methods. The first method involves using the Wide-field Infrared Survey Explorer (WISE) colours \citep{wright_2010}, based on the $W1$ and $W2$ bands, which correspond to 3.5 $\mu$m and 4.6 $\mu$m, respectively \citep{assef_2018}. The MaNGA AGN catalogue employed the 75\% reliability criteria, defined as $W1 - W2 > 0.486 \exp{ \left[0.092 (W2 - 13.07)^{2} \right]}$ and $W2 > 13.07$, or $W1 - W2 > 0.486$ and $W2 \leq 13.07$ \citep{assef_2018,comerford_2020}. The second method is based on X-ray sources from the Swift observatory's Burst Alert Telescope (BAT) all-sky survey. The BAT catalogue has identified AGN and cross-matched them with SDSS \citep{oh_2018}, which are included in the MaNGA AGN catalogue \citep{comerford_2020}. The third method utilized radio observations. \citet{best_2012} have created an AGN catalogue based on the NRAO Very Large Array Sky Survey (NVSS, \citealp{condon_1998}) and the Faint Images of the Radio Sky at Twenty centimeters survey (FIRST, \citealp{becker_1995}). This catalogue was then cross-matched against MaNGA and added in the MaNGA AGN catalogue \citep{comerford_2020}. Finally, AGN were also identified in the MaNGA AGN catalogue using broad H$\alpha$ emission lines in spectra from SDSS \citep{oh_2015}. They identified AGN based on a few thresholds. Firstly, the full-width at half-maximum (FWHM) of the broad H$\alpha$ component has to be higher than 800 km s$^{-1}$ and the amplitude over noise (A/N) ratio of the broad H$\alpha$ component must be larger than 3. Additionally, the ratio of the area of the broad H$\alpha$ component to the noise level of the continuum (more specifically: the measurement uncertainty around [$\textrm{N}\scriptstyle\mathrm{II}$] $\lambda 6584$) must be greater than 2 \citep{oh_2015}.

The AGN identified with the broad emission lines are expected to affect the results presented here the most, as they will interfere with the H$\alpha$ measurement. This method identified \todo{14} out of \todo{2,125} galaxies (\todo{0.7}\%) in the GZ DESI-MaNGA sample as an AGN. A total of \todo{106} galaxies in the GZ DESI-MaNGA sample are identified as an AGN in the MaNGA AGN catalogue using any of the methods described above, implying an overall AGN fraction of \todo{5.0}\%. However, the MaNGA AGN catalogue only covers 6,261 of the 11,273 MaNGA targets \citep{comerford_2020}. The total estimated AGN fraction of the GZ DESI-MaNGA sample is \todo{1.2}\% when considering only AGN identified by the broad emission lines, assuming a similar AGN fraction for the remaining galaxies. When including AGN identified by any of the methods discussed above, the total estimated AGN fraction becomes \todo{9.0}\%. To ensure that the presence of AGN did not introduce any bias into the results presented here, we excluded the \todo{106} galaxies identified as AGN by the MaNGA AGN catalogue using any method from the sample. We found that all the results presented here remained qualitatively the same. We can therefore conclude that the results in this paper are not significantly biased by the presence of AGN.


\section{Results}


\subsection{Strong and weak bars}
\label{sec:r_profile_sb_wb}

Observations show that the bar fraction is higher among quiescent galaxies \citep{masters_2011,masters_2012,cervantessodi_2017,vera_2016}, which implies that bars are involved in the quenching process. To understand how a bar affects its host, studies have looked at the star formation in different regions of the bar, such as the centre and arms of the bar (e.g. see \citealp{watanabe_2011,haywood_2016, fraser_mckelvie_2020,lin_2020,maeda_2020b}). However, it remains unclear whether these regions are affected similarly in both weak and strong bars. This is the topic of the first part of this paper, where we use resolved star formation indicators obtained through the MaNGA pipeline to look at the differences in star formation between strong and weak bars in detail.

\subsubsection{EW[H$\alpha$] along strong and weak bars}
\label{sec:result_sbwb_ewha}

\begin{figure*}
    \centering
    \includegraphics[width=\textwidth]{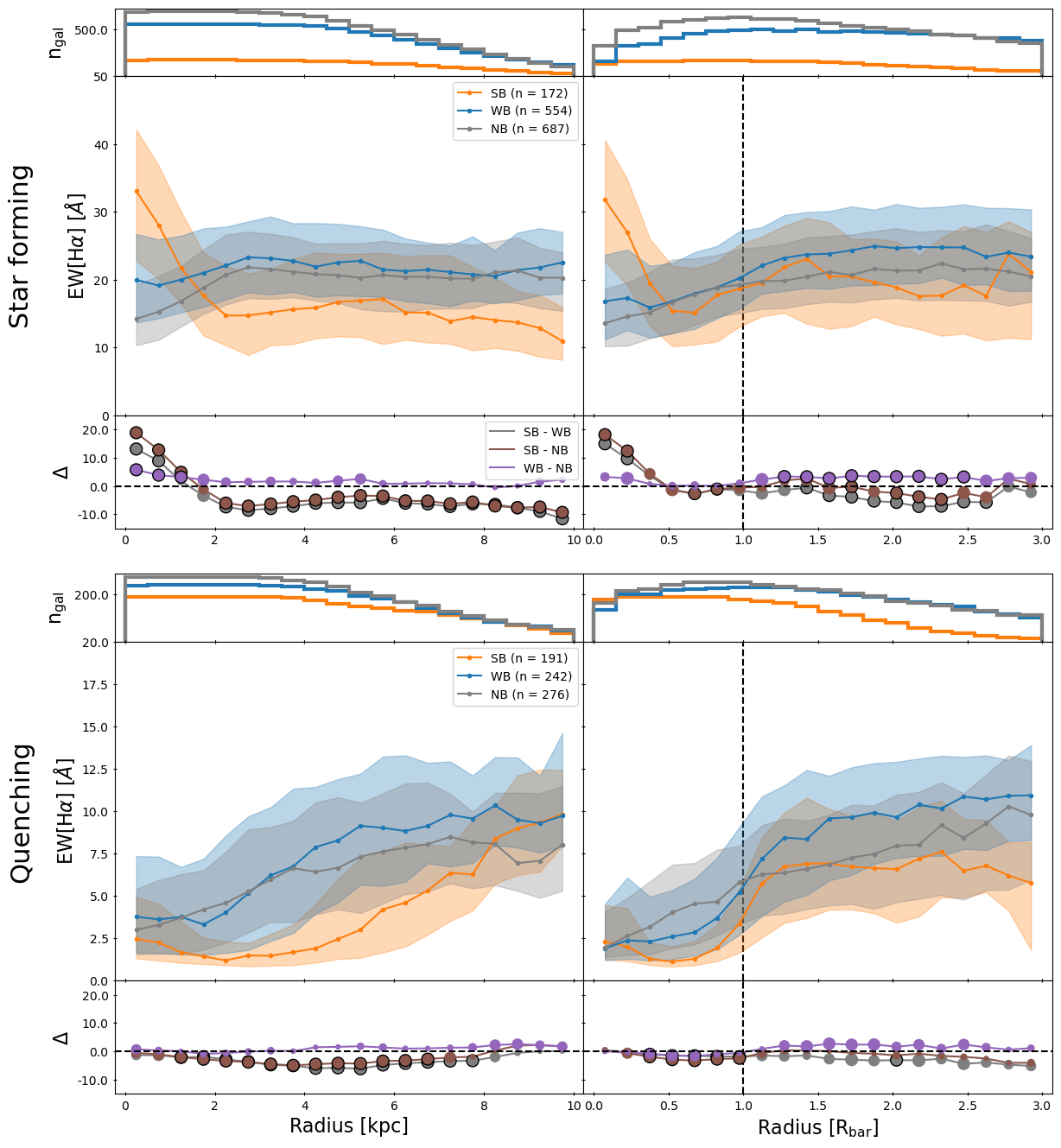}
    \caption{The radial profiles of EW[H$\alpha$] are shown for SF galaxies (top row) and quenching galaxies (bottom row) for strongly barred (orange), weakly barred (blue) and unbarred galaxies (grey). The radial profiles are shown in kpc (left column) and normalised against the bar radius (right column). The profiles for unbarred galaxies are normalised to the bar radius by assigning a random bar radius to every unbarred galaxy, which is drawn from a log-normal distribution fitted to all the barred galaxies in the sample. The sample size, n, for every sample is indicated in the legend. The solid line is the median in every bin, while the shaded regions bound the 33$^{\rm rd}$ and 66$^{\rm th}$ percentiles. The dashed line at $R_{\textrm{bar}}$ = 1 denotes the end of the bar. The width of the bins in the left column is \todo{0.5} kpc and \todo{0.15 $R_{\textrm{bar}}$} in the right column. The top part of each panel shows the number of galaxies that have spaxels in each radial bin. The bottom part of each panel shows the difference between each radial profile in every bin, where the size of the point represents the significance of the difference after comparing the two populations with an Anderson-Darling test. The smallest sizes represent a significant difference of less than 1$\sigma$, while the largest sizes represent $>$3$\sigma$ and are additionally outlined in black. Strongly barred SF galaxies have higher values for EW[H$\alpha$] in the centre and bar-end, while having less EW[H$\alpha$] in the arms of the bar. The profiles of weak and unbarred SF galaxies are very similar to each other.}
    \label{fig:r_profile_ewha}
\end{figure*}

We generate radial profiles for EW[H$\alpha$] and D$_{\rm n}$4000 obtained from MaNGA in order to probe the effect that a bar has on different regions of a galaxy. These radial profiles are constructed for all the galaxies in the GZ DESI-MaNGA sample. Figure \ref{fig:r_profile_ewha} shows the radial profiles of the Gaussian-fitted equivalent width measurement of H$\alpha$ (EW[H$\alpha$]) for each bar type: strong, weak, and unbarred. The galaxies are categorised as SF or quenching based on Equation \ref{eq:sfr_mass_split} shown in Section \ref{sec:sample_selection}. The top row shows the profiles for SF galaxies, while the bottom row shows the profiles for quenching galaxies. The profiles are shown in terms of distance to the centre of the galaxy (in kpc, left column) and relative to the bar radius (right column). However, it is important to realise that a distance of e.g. $2 R_{\textrm{bar}}$ is farther out in the disc of the galaxy for a strong bar than for a weak bar, as strong bars are typically longer than weak bars \citep{devaucouleurs_1959, devaucouleurs_1963,geron_2021}. The top part of each panel shows the number of galaxies that have spaxels in each radial bin. The bottom part of every panel shows the difference between the radial profiles of any two samples. The significance of the difference in every bin is shown by the size of the point, with the largest sizes representing a significant difference greater than 3$\sigma$ after comparing the two populations with an Anderson-Darling test. Conversely, the smallest sizes represent $<$1$\sigma$. Bins with a significant difference exceeding 3$\sigma$ are additionally outlined in black.

The EW[H$\alpha$] radial profile of strongly barred SF galaxies (orange, top row of Figure \ref{fig:r_profile_ewha}) peaks in the central bins, after which it reaches a minimum in the arms of the bar. Interestingly, the profile rises again, reaching another peak beyond the bar-end region (at $R \approx 1.2-1.5 R_{\textrm{bar}}$), before decreasing again at higher radii. This profile suggests a higher SFR in the centre and beyond the bar-end, as EW[H$\alpha$] serves as a proxy for sSFR, while it is suppressed in the arms of the bar. The EW[H$\alpha$] profiles of weakly barred SF galaxies and unbarred SF galaxies closely resemble each other. Both profiles are flat in the outskirts (at R $\gtrsim $ 5 kpc or $\gtrsim$ 1.5 $R_{\textrm{bar}}$) and have lower median values for EW[H$\alpha$] closer to the centre. Additionally, it is interesting to note that the EW[H$\alpha$] is lower for strongly barred SF galaxies compared to weakly barred and unbarred SF galaxies in the outskirts of the galaxy.

The bottom half of Figure \ref{fig:r_profile_ewha} shows the radial profiles for quenching galaxies with strong bars, weak bars, and no bars. Strongly barred quenching galaxies have consistently lower values for EW[H$\alpha$], while weakly and unbarred galaxies have relatively similar values to each other. All profiles have higher values for EW[H$\alpha$] in the outskirts of the galaxy, which decrease towards the centre, which is the opposite trend to what we saw for strongly barred SF galaxies. Interestingly, for both weak and strong bars, this decrease starts around the bar-end ($R \approx 1.0-1.2 R_{\textrm{bar}}$), although it is more distinct for strong bars. This decrease at the bar-end region is not found in the profiles of the unbarred galaxies normalised to the bar radius. However, it is important to keep in mind that the profiles normalised to the bar radius for unbarred galaxies are not physical. They are constructed by assigning a random bar radius to every unbarred galaxy, which is drawn from a log-normal distribution fitted to all the barred galaxies in the same stellar mass bin to the unbarred galaxy (see Section \ref{sec:radius_profile_method} for more details).

\subsubsection{D$_{\rm n}$4000 along strong and weak bars}

\begin{figure*}
  \centering
    \includegraphics[width=\textwidth]{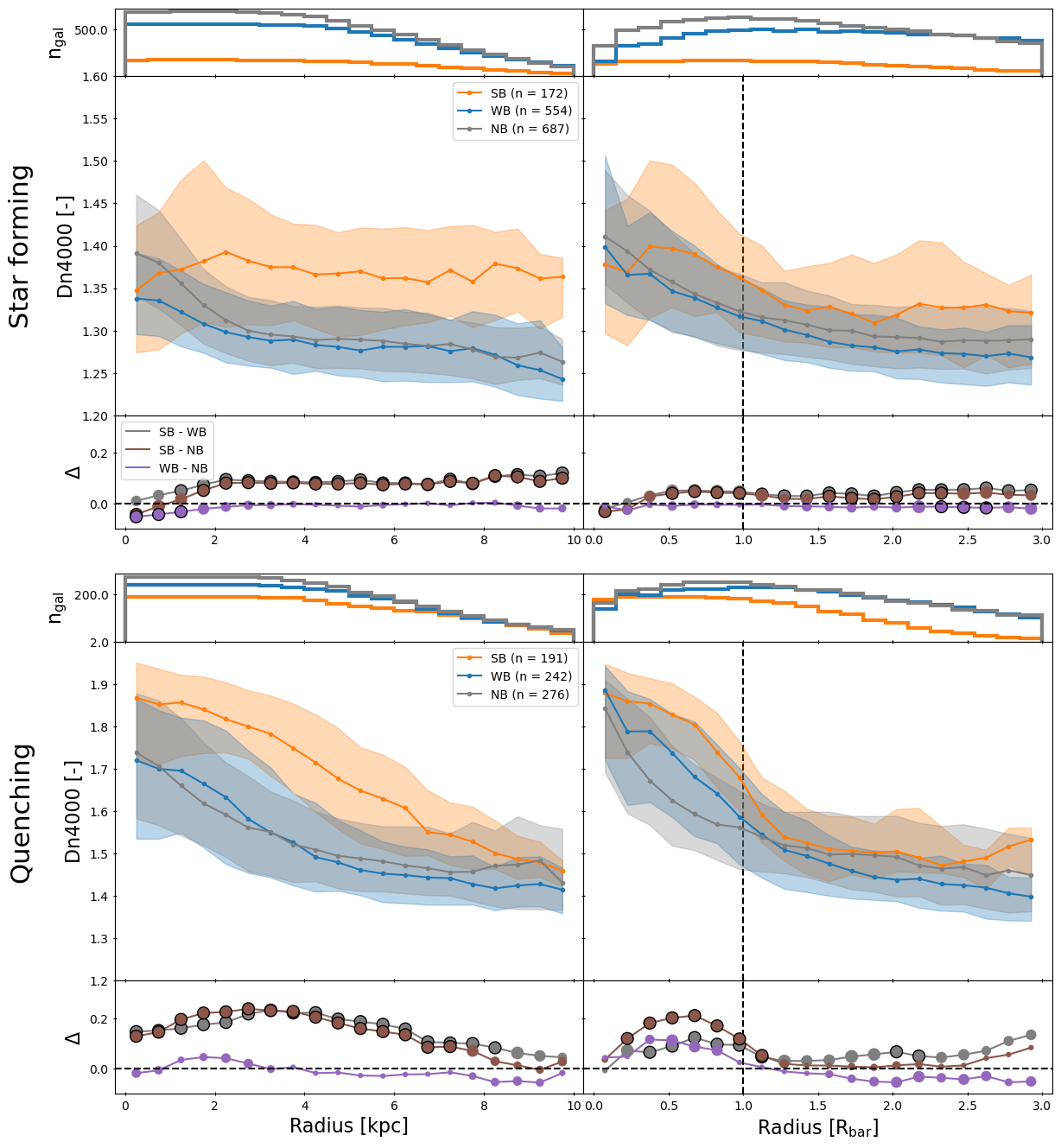}
    \caption{The radial profiles of D$_{\rm n}$4000 are shown for SF galaxies (top row) and quenching galaxies (bottom row) for strongly barred (orange), weakly barred (blue) and unbarred galaxies (grey). The radial profiles are shown in kpc (left column) and normalised against the bar radius (right column). The profiles for unbarred galaxies are normalised to the bar radius by assigning a random bar radius to every unbarred galaxy, which is drawn from a log-normal distribution fitted to all the barred galaxies in the sample. The sample size, n, for every sample is indicated in the legend. The solid line is the median in every bin, while the shaded regions bound the 33$^{\rm rd}$ and 66$^{\rm th}$ percentiles. The dashed line at $R_{\textrm{bar}}$ = 1 denotes the end of the bar. The width of the bins in the left column is \todo{0.5} kpc and \todo{0.15 $R_{\textrm{bar}}$} in the right column. The top part of each panel shows the number of galaxies that have spaxels in each radial bin. The bottom part of each panel shows the difference between each radial profile in every bin, where the size of the point represents the significance of the difference after comparing the two populations with an Anderson-Darling test. The smallest sizes represent a significant difference of less than 1$\sigma$, while the largest sizes represent $>$3$\sigma$ and are additionally outlined in black. Strongly barred SF galaxies have lower values for D$_{\rm n}$4000 in the centre, while having higher values for D$_{\rm n}$4000 in the arms of the bar. The profiles of weak and unbarred SF galaxies are very similar to each other.}
    \label{fig:r_profile_dn4000}
\end{figure*}

Figure \ref{fig:r_profile_dn4000} shows the radial profiles for D$_{\rm n}$4000. The average profile for strongly barred SF galaxies is low in the centre of the galaxy, reaches a maximum in the middle of the arms of the bar ($R \approx 0.5 R_{\textrm{bar}}$), followed by a decline and flatting beyond the bar-end region ($R \approx 1.2-1.5 R_{\textrm{bar}}$). This profile is in contrast from the profiles observed in weakly barred and unbarred galaxies. Here, D$_{\rm n}$4000 is highest in the centre and decreases monotonically with radius. On average, strongly barred galaxies have the highest values for D$_{\rm n}$4000, followed by weakly barred and unbarred SF galaxies, which have very similar values to each other. These results suggest that the stellar population is oldest in the arms of the bar of strongly barred SF galaxies and that they have younger stellar populations in the centre and beyond the bar-end. These results are consistent with the results obtained from the EW[H$\alpha$] radial profiles. These results confirm that strong bars can impact stellar populations, implying that strong bars are a long-lived phenomena. This is not the case for weakly barred galaxies, whose D$_{\rm n}$4000 profiles closely resemble those of unbarred galaxies. These results are discussed in more detail in Section \ref{sec:disc_sbwb_lifetime}.

Among quenching galaxies (bottom row of Figure \ref{fig:r_profile_dn4000}), the profiles of the three samples exhibit similar trends: highest in the centre and decreasing monotonically with radius. However, for strongly barred galaxies, this decrease happens around the bar-end region, whereas it is much more gradual for weakly barred and unbarred galaxies. In the outskirts of the galaxy, the value for D$_{\rm n}$4000 is comparable for all three samples. Strongly barred quenching galaxies have higher values for D$_{\rm n}$4000 overall, while weakly and unbarred galaxies have similar values to each other.

To summarise: \textbf{strongly barred SF galaxies have increased star formation in the centre and bar-end region, while suppressing star formation in the arms of the bar. In contrast, the profiles of weakly barred and unbarred galaxies are very similar to each other}. Weakly and unbarred galaxies have lower star formation in the centre, which increases with radius. These results are further discussed in greater detail and compared to the literature in Section \ref{sec:disc_sbwb_barregions}.

\subsubsection{Effect of stellar mass}
\label{app:sbwb_mass}

Unbarred, weakly barred and strongly barred galaxies typically have different stellar masses \citep{geron_2021}. To investigate how stellar mass impacts the results of this paper, we also generated the radial profiles for all SF galaxies across three different mass bins: low mass (\todo{$M_{*} < 10^{10.2} M_{\odot}$}), intermediate mass (\todo{$10^{10.2} M_{\odot} < M_{*} < 10^{10.6} M_{\odot}$}) and high mass (\todo{$M_{*} > 10^{10.6} M_{\odot}$}). These thresholds were selected so that an equal number of galaxies are in each bin (\todo{471} galaxies in each). 

\begin{figure*}
  \centering
    \includegraphics[width=0.8\textwidth]{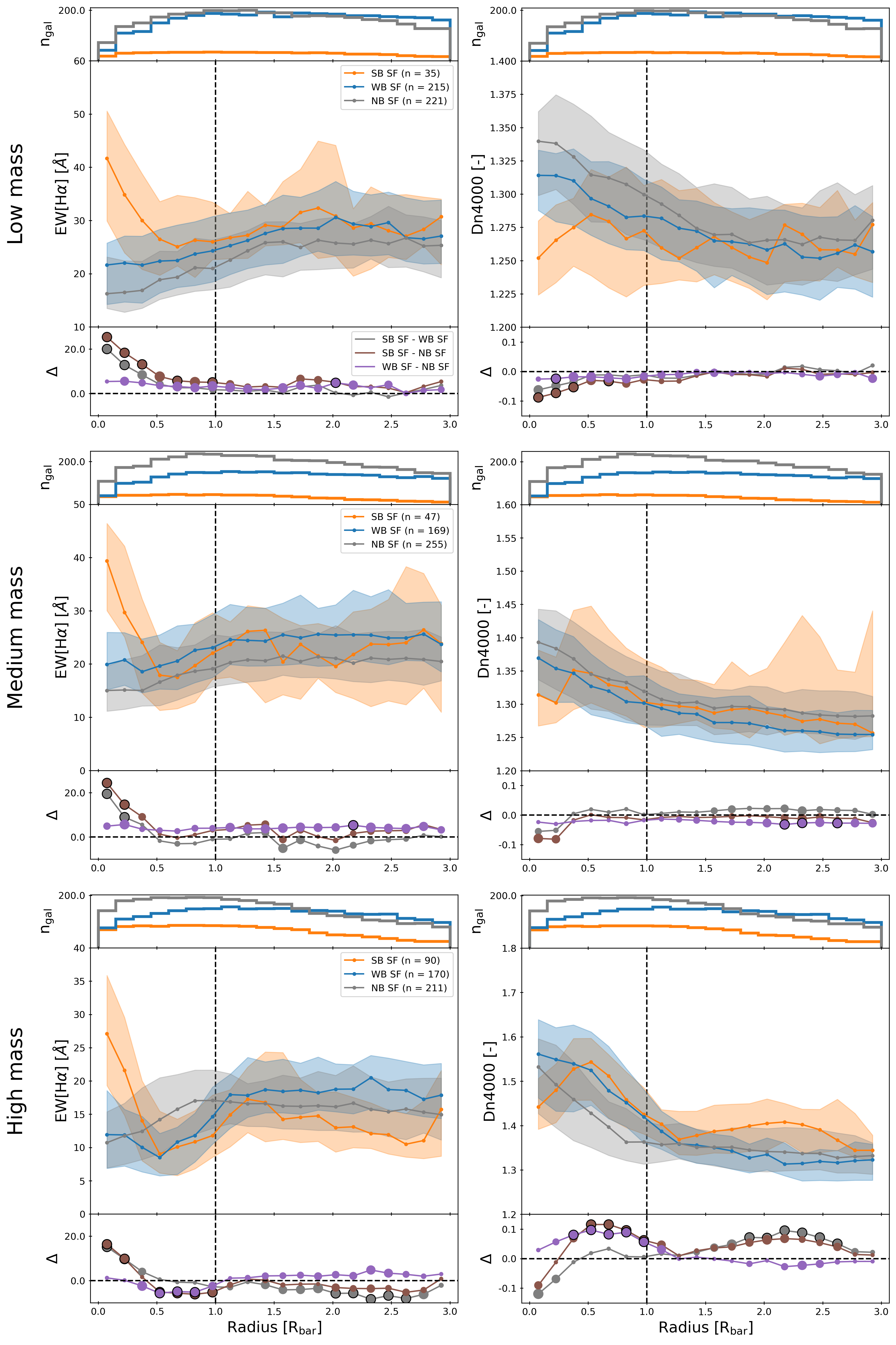}
    \caption{The radial profiles of EW[H$\alpha$] (left column) and D$_{\rm n}$4000 (right column) are shown for SF galaxies that are strongly barred (orange), weakly barred (blue) and unbarred (grey). These profiles are constructed similarly to those in Figures \ref{fig:r_profile_ewha} and \ref{fig:r_profile_dn4000}. However, the profiles here are shown in three different mass bins: low mass (\todo{$M_{*} < 10^{10.2} M_{\odot}$}, top row), intermediate mass (\todo{$10^{10.2} M_{\odot} < M_{*} < 10^{10.6} M_{\odot}$}, middle row) and high mass (\todo{$M_{*} > 10^{10.6} M_{\odot}$} bottom row). The profiles for intermediate and high mass galaxies are similar to those found in Figures \ref{fig:r_profile_ewha} and \ref{fig:r_profile_dn4000}. This is not true for the low mass galaxies.}
    \label{fig:r_profile_mass}
\end{figure*}

The left column of Figure \ref{fig:r_profile_mass} shows the radial profiles of EW[H$\alpha$] for the three mass bins. The central peak of EW[H$\alpha$] for strongly barred galaxies is found in all mass bins. However, the suppression of EW[H$\alpha$] and the subsequent second peak beyond the bar-end region ($R \approx 1.2-1.5 R_{\textrm{bar}}$) among strongly barred SF galaxies are only observed in the intermediate and high mass bins, not in the low mass bin. Interestingly, the EW[H$\alpha$] of the weakly barred galaxies also increases around the bar-end in the high mass bin. A similar result is shown in in right column of Figure \ref{fig:r_profile_mass}, where the radial profiles of D$_{\rm n}$4000 for SF galaxies in the different mass bins are shown. The increase of D$_{\rm n}$4000 in the arms of strong bars is much more prominent in the intermediate mass and high mass bins, compared to the low mass bin. The profiles of the weakly barred SF galaxies and unbarred SF galaxies are similar across all mass bins in both the EW[H$\alpha$] and D$_{\rm n}$4000 radial profiles.

These results show that \textbf{the conclusions drawn in this paper regarding bar strength hold true for intermediate mass and high mass galaxies, but not for low mass galaxies}. This is in agreement with the results found in \citet{geron_2021}, who showed that strong bars are most efficient at facilitating quenching at higher stellar masses. Other studies have found that more massive galaxies have longer and stronger bars \citep{aguerri_2009,erwin_2019,fraser_mckelvie_2020,kim_2021}. This implies that the strongest bars, which have the most effect on their host, are predominantly found in the highest mass bin.

\subsubsection{Focus on bar-end region}
\label{sec:result_sbwb_barend}

\begin{figure*}
  \centering
    \includegraphics[width=\textwidth]{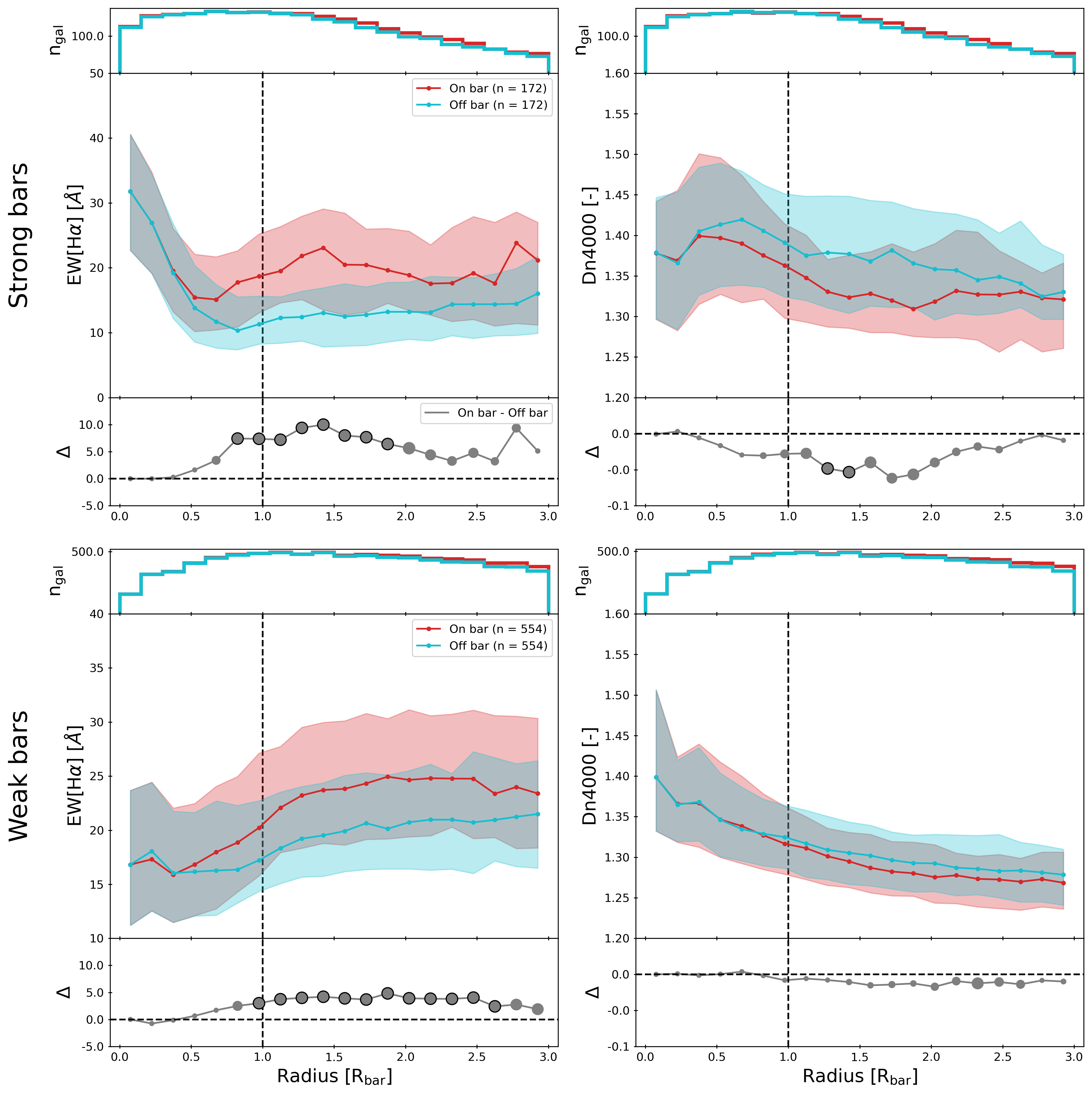}
    \caption{The effect of the bar-end region is probed in more detail in SF galaxies by creating radial profiles with apertures placed on the bar (parallel to the PA of the bar, shown in red) and off the bar (perpendicular to the PA of the bar, shown in light blue). This is done for the EW[H$\alpha$] profiles (left column) and the D$_{\rm n}$4000 profiles (right column). The width of the bins is \todo{0.15 $R_{\textrm{bar}}$}. The difference is much greater for strongly barred SF galaxies (top row) than for weakly barred SF galaxies (bottom row). The difference in EW[H$\alpha$] and D$_{\rm n}$4000 in the bar-end is bigger for strongly barred galaxies than weakly barred SF galaxies.}
    \label{fig:r_profile_barend_sbwb}
\end{figure*}

The bar-end region is of particular interest. A local peak of EW[H$\alpha$] is found here for strongly barred SF galaxies, but not for weakly barred SF galaxies. Moreover, the bar-end region appears to be a turn-over point in the EW[H$\alpha$] and D$_{\rm n}$4000 profiles for strongly barred galaxies. To study this region in more detail and compare it to corresponding off-bar regions, we construct radial profiles with the aperture placed parallel and perpendicular to the bar (the rest of the procedure remains identical, see Section \ref{sec:radius_profile_method} for more details). The profiles created with the aperture placed parallel to the bar (as shown in Figures \ref{fig:r_profile_ewha}, \ref{fig:r_profile_dn4000} and \ref{fig:r_profile_mass}) provide information specifically about the barred region, such as the bar-end. In contrast, profiles created with the aperture placed perpendicular to the bar probe the off-bar regions. Comparing the profiles generated with these two apertures will show what the impact of the bar is (e.g. in the bar-end region) with respect to the rest of the galaxy. However, do note that the central region of the profiles should be similar, given that the parallel and perpendicular apertures overlap in the \todo{3} arcsec centre.

This is done in Figure \ref{fig:r_profile_barend_sbwb} for all strongly barred SF galaxies (top row) and weakly barred SF galaxies (bottom row) for both EW[H$\alpha$] (left column) and D$_{\rm n}$4000 (right column). The median difference in EW[H$\alpha$] beyond the bar-end region ($R \approx 1.2-1.5 R_{\textrm{bar}}$) is $\sim$10~\angstrom{} for strongly barred SF galaxies, while this is only $\sim$4~\angstrom{} for weakly barred SF galaxies, though it is $>$3$\sigma$ significantly different for both strongly and weakly barred SF galaxies. Additionally, the profile with the aperture aligned with the bar clearly shows the peak beyond the bar-end region, which this is not observed in the profile with the aperture placed perpendicular to the bar. The difference in D$_{\rm n}$4000 at the bar-end region for strongly barred SF galaxies is $\sim$0.05 and it is $>$3$\sigma$ significant. Meanwhile, the difference in D$_{\rm n}$4000 is only $\sim$0.01 for weakly barred SF galaxies and this difference not significant ($<$2$\sigma$) in any bin.

Thus, the difference in EW[H$\alpha$] and D$_{\rm n}$4000 around the bar-end region is significantly higher for strongly barred SF galaxies compared to weakly barred SF galaxies. This suggests that there is \textbf{more star formation in the bar-end region of strongly barred SF galaxies compared to the bar-end region of weakly barred SF galaxies}.


\subsection{Fast and slow bars}
\label{sec:r_profile_fb_sb}

As explained in Section \ref{sec:tw_method}, bars are kinematically classified as either fast or slow based on $\mathcal{R}$, the ratio of the corotation radius to the bar radius. In this section, we will investigate whether there are significant differences between fast and slow bars in terms of their EW[H$\alpha$] and D$_{\rm n}$4000 profiles. This will probe whether the kinematics of bars have a measurable impact on the star formation of their hosts. Futhermore, we will examine whether the kinematics of the bar is influenced by the bar strength, i.e. whether fast strong bars have different effects on their hosts compared to slow strong bars.

\subsubsection{EW[H$\alpha$] and D$_{\rm n}$4000 along fast and slow bars}
\label{sec:result_fbsb_ewha}

\begin{figure*}
  \centering
    \includegraphics[width=\textwidth]{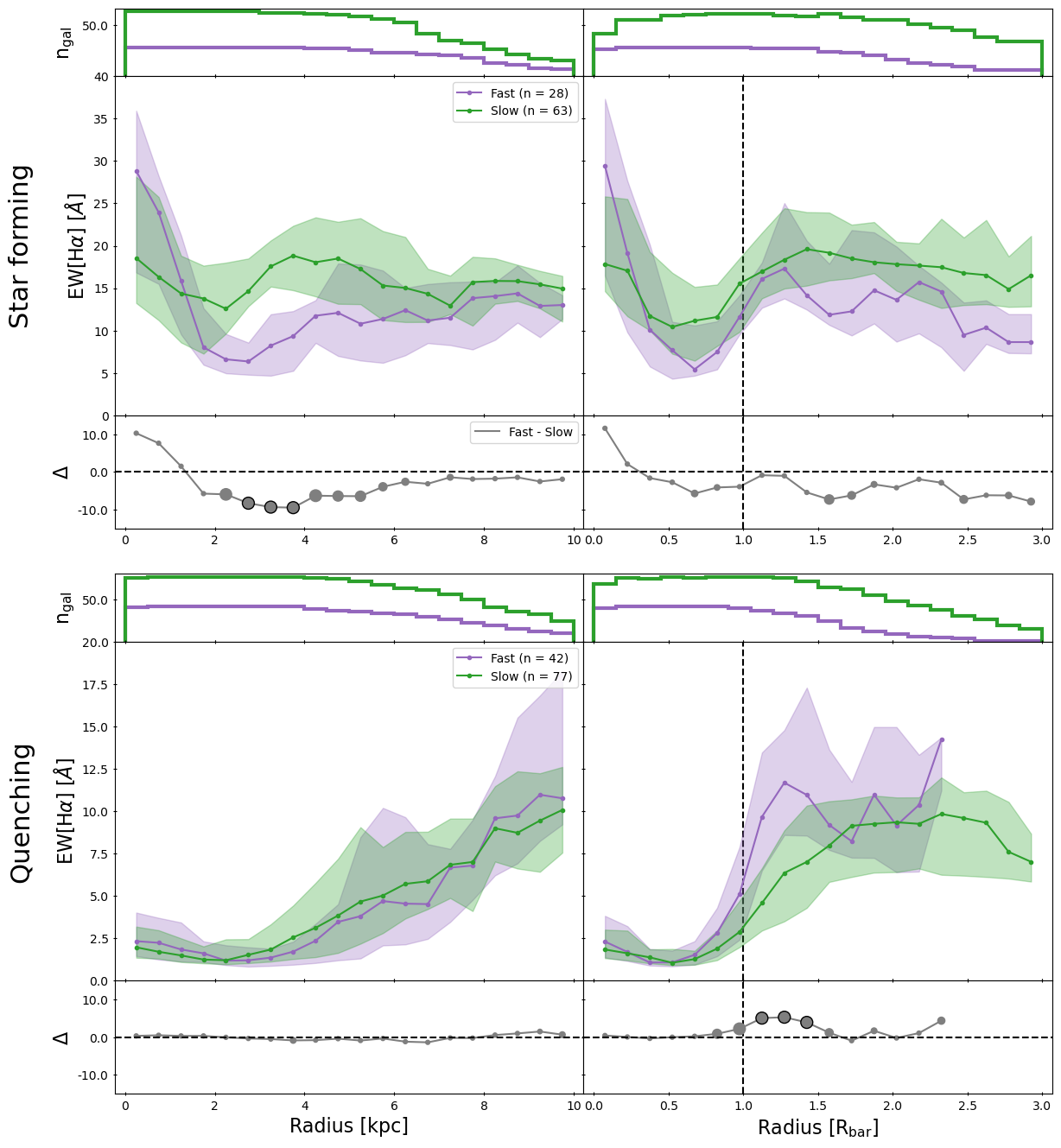}
    \caption{The radial profiles of EW[H$\alpha$] are shown for SF galaxies (top row) and quenching galaxies (bottom row) for fast bars (purple) and slow bars (green). The radial profiles are shown in kpc (left column) and normalised against the bar radius (right column). The sample size, n, for every sample is indicated in the legend. The solid line is the median in every bin, while the shaded regions bound the 33$^{\rm rd}$ and 66$^{\rm th}$ percentiles. The dashed line at $R_{\textrm{bar}}$ = 1 denotes the end of the bar. The width of the bins in the left column is \todo{0.5} kpc and \todo{0.15 $R_{\textrm{bar}}$} in the right column. The top part of each panel shows the number of galaxies that have spaxels in each radial bin. The bottom part of each panel shows the difference between each radial profile in every bin, where the size of the point represents the significance of the difference after comparing the two populations with an Anderson-Darling test. The smallest sizes represent a significant difference of less than 1$\sigma$, while the largest sizes represent $>$3$\sigma$ and are additionally outlined in black. Slow bars in SF galaxies have higher values for EW[H$\alpha$] in the intermediate radius range of the galaxy than fast bars.}
    \label{fig:r_profile_ewha_fbsb}
\end{figure*}

The radial profiles of EW[H$\alpha$] for fast and slow bars in SF and quenching galaxies are shown in Figure \ref{fig:r_profile_ewha_fbsb}, which is constructed similarly as before. The top-left panel clearly shows that slow bars have higher EW[H$\alpha$] than fast bars in SF galaxies, especially in the intermediate radius range of the galaxy ($\sim$\todo{2-4} kpc). The top-right panel shows that the median profile is slightly higher for slow bars than for fast bars in SF galaxies when expressing distance normalised to the bar radius, although the difference remains less than 3$\sigma$. Both slow and fast bars have higher EW[H$\alpha$] in the centre and bar-end compared to the arms of the bar. The profiles for quenching galaxies are relatively similar between slow and fast bars, except around the bar-end, where the EW[H$\alpha$] profile is steeper for fast bars and fast bars have significantly ($>$3$\sigma$) higher values for EW[H$\alpha$]. The rest of the EW[H$\alpha$] profile rises slowly with increasing distance from the galaxy.

\begin{figure*}
  \centering
    \includegraphics[width=\textwidth]{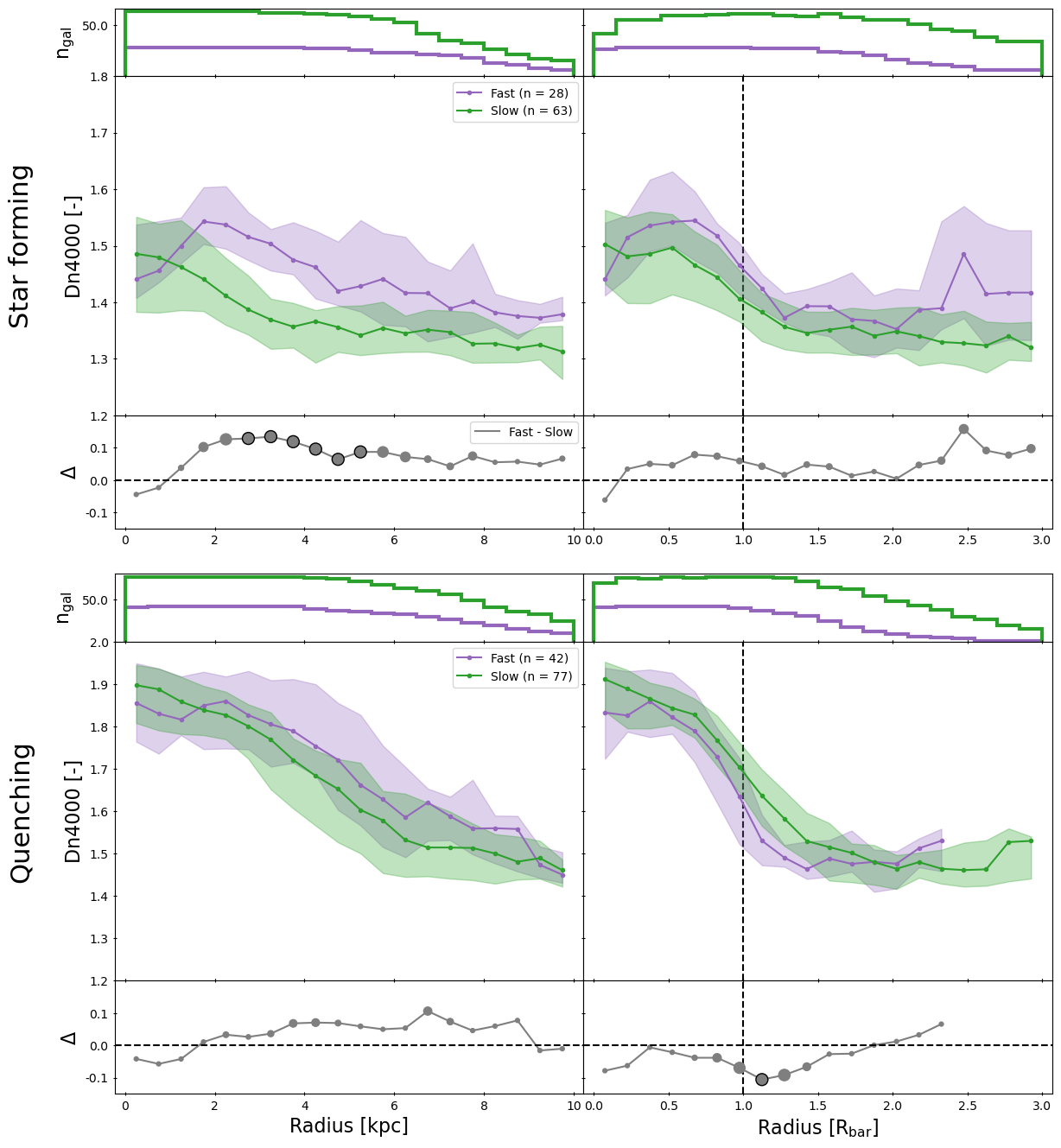}
    \caption{The radial profiles of D$_{\rm n}$4000 are shown for SF galaxies (top row) and quenching galaxies (bottom row) for fast bars (purple) and slow bars (green). The radial profiles are shown in kpc (left column) and normalised against the bar radius (right column). The sample size, n, for every sample is indicated in the legend. The solid line is the median in every bin, while the shaded regions bound the 33$^{\rm rd}$ and 66$^{\rm th}$ percentiles. The dashed line at $R_{\textrm{bar}}$ = 1 denotes the end of the bar. The width of the bins in the left column is \todo{0.5} kpc and \todo{0.15 $R_{\textrm{bar}}$} in the right column. The top part of each panel shows the number of galaxies that have spaxels in each radial bin. The bottom part of each panel shows the difference between each radial profile in every bin, where the size of the point represents the significance of the difference after comparing the two populations with an Anderson-Darling test. The smallest sizes represent a significant difference of less than 1$\sigma$, while the largest sizes represent $>$3$\sigma$ and are additionally outlined in black. Slow bars in SF galaxies have lower values for D$_{\rm n}$4000 in the intermediate radius range of the galaxy than fast bars.}
    \label{fig:r_profile_dn4000_fbsb}
\end{figure*}

Similar conclusions are drawn for the D$_{\rm n}$4000 profiles, which are shown in Figure \ref{fig:r_profile_dn4000_fbsb}. Slow bars in SF galaxies have significantly lower values for D$_{\rm n}$4000 in the intermediate radius range of the galaxy ($\sim$3-5 kpc) than fast bars. The average profile for slow bars in SF galaxies decreases monotonically with radius, while the profile for fast bars has a bump in the arms of the bar. Again, the profiles of slow and fast bars in quenching galaxies are relatively similar, except at the bar-end region.

These results suggest that \textbf{there is more star formation along slow bars compared to fast bars in SF galaxies}, as we find higher values for EW[H$\alpha$] and lower values for D$_{\rm n}$4000 in slow bars than in fast bars.

\subsubsection{Effect of bar strength}

The results presented in Section \ref{sec:result_fbsb_ewha} suggest that there is more star formation along slow bars compared to fast bars in SF galaxies. At the same time, we showed in Section \ref{sec:r_profile_sb_wb} that there is more star formation in strongly barred galaxies than in weakly barred galaxies. In this section, we will address whether there is a combined effect between these two independent results.

\begin{figure*}
  \centering
    \includegraphics[width=\textwidth]{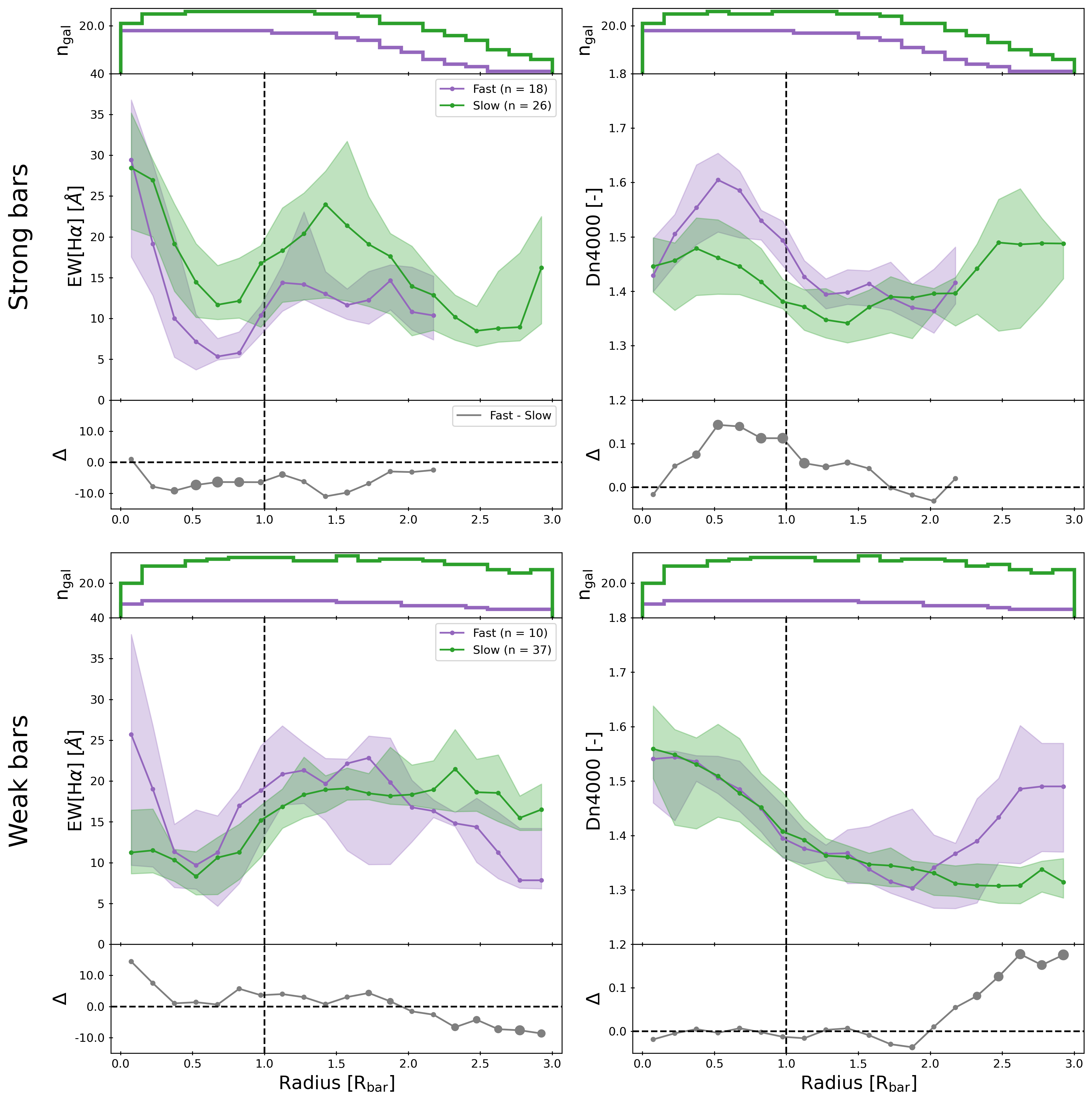}
    \caption{The radial profiles of EW[H$\alpha$] (left column) and D$_{\rm n}$4000 (right column) are shown for strongly barred SF galaxies (top row) and weakly barred SF galaxies galaxies (bottom row). The sample is additionally divided into fast bars (purple) and slow bars (green). The radial profiles are shown normalised against the bar radius (right column). The sample size, n, for every sample is indicated in the legend. The solid line is the median in every bin, while the shaded regions bound the 33$^{\rm rd}$ and 66$^{\rm th}$ percentiles. The dashed line at $R_{\textrm{bar}}$ = 1 denotes the end of the bar. The width of the bins in is \todo{0.15 $R_{\textrm{bar}}$}. The top part of each panel shows the number of galaxies that have spaxels in each radial bin. The bottom part of each panel shows the difference between each radial profile in every bin, where the size of the point represents the significance of the difference after comparing the two populations with an Anderson-Darling test. The smallest sizes represent a significant difference of less than 1$\sigma$, while the largest sizes represent $>$3$\sigma$ and are additionally outlined in black. It is clear that a slow bar will affect its host more if it is also strong. The profiles of fast and slow weak bars are very similar to each other.}
    \label{fig:r_profile_fbsb_sbwb_sf}
\end{figure*}

The effect of bar strength on the profiles of fast and slow bars is studied in Figure \ref{fig:r_profile_fbsb_sbwb_sf}, which shows the radial profiles for EW[H$\alpha$] (left column) and D$_{\rm n}$4000 (right column) for strongly barred SF galaxies (top row) and weakly barred SF galaxies (bottom row). The median profile of EW[H$\alpha$] for slow strong bars is always higher than that of fast strong bars, as shown in the top-left panel of Figure \ref{fig:r_profile_fbsb_sbwb_sf}, although the significance of the difference is always $<$3$\sigma$. These profiles are reminiscent of the profiles previously shown in Figure \ref{fig:r_profile_ewha} for strong bars: high EW[H$\alpha$] in the centre, lower values in the arm of the bar and a second peak beyond the bar-end. However, the increase of EW[H$\alpha$] beyond the bar-end region is more pronounced in slow strong bars than in fast strong bars. The median EW[H$\alpha$] is also higher in the arms of slow strong bars than in the arms of fast strong bars. The EW[H$\alpha$] profiles for slow weak bars and fast weak bars, shown in the bottom-left panel, are similar to each other ($<$1$\sigma$ at all radii smaller than 2$R_{\rm bar}$).

The top-right panel shows that the median values for D$_{\rm n}$4000 are always lower for slow strong bars than for fast strong bars, which is consistent with the EW[H$\alpha$] profiles. Fast strong bars have a very distinct peak of D$_{\rm n}$4000 in the arms of the bar, which is not as apparent in slow strong bars. This suggests that galaxies with slow strong bars might have younger stellar populations in the arms of the bar due to more recent star formation, presumably triggered by the presence of the slow strong bar. However, these differences are again not significantly different ($<$3$\sigma$), possibly due to the very low sample size. The bottom-right panel shows that the D$_{\rm n}$4000 profiles for fast weak and slow weak bars are very similar to each other ($<$1$\sigma$ at all radii smaller than 2$R_{\rm bar}$).

Despite the limited sample sizes used in this analysis, bars strength seems to have an effect on the profiles of fast and slow bars. While not statistically significant ($<$3$\sigma$), these results suggest that there is \textbf{more star formation at all radii in slow strong bars, compared to fast strong bars. In contrast, the profiles of fast and slow weak bars are very similar to each other}. Increasing the sample size in future studies will be crucial to clarify these issues, as discussed in more detail in Section \ref{sec:disc_fbsb_barstrength}. The physical interpretation of this observation will be addressed in the following two subsections.

\subsubsection{Global or local effect?}
\label{sec:result_fbsb_localeffect}

\begin{figure}
    \includegraphics[width=\columnwidth]{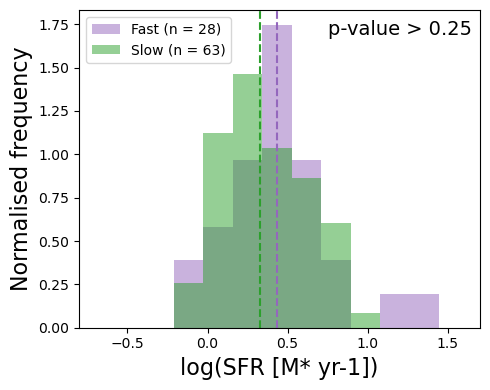}
    \caption{The difference in global star formation rate (obtained from Pipe3D) between fast bars in SF galaxies (purple) and slow bars in SF galaxies (green). The p-value of a two-sample Anderson-Darling test is shown in the top-right corner, with the null hypothesis being that the two samples are drawn from the same population. The global star formation rate is not significantly different between fast and slow bars.}
    \label{fig:sfr_fbsb}
\end{figure}

We have shown that slow bars have higher EW[H$\alpha$] than fast bars in SF galaxies, which implies more star formation in slow bars than fast bars. Interestingly, Figure \ref{fig:sfr_fbsb} shows that the global SFR (obtained from Pipe3D, \citealp{sanchez_2016,sanchez_2016b,sanchez_2018}) is not significantly different between SF galaxies with fast and slow bars (the p-value of an Anderson-Darling test between the SFRs of fast and slow bars is $>$0.25). These observations initially seem contradictory to each other. However, the radial profiles are constructed using a aperture with a width of \todo{3} arcsec oriented along the position angle (PA) of the bar. This means that these results suggest that the increase in EW[H$\alpha$] observed in slow bars only affects the bar region locally, without a significant impact on the global star formation of the galaxy. In other words, a slow bar seems to concentrate its SF along the bar, without increasing the global star formation.

\begin{figure*}
  \centering
    \includegraphics[width=\textwidth]{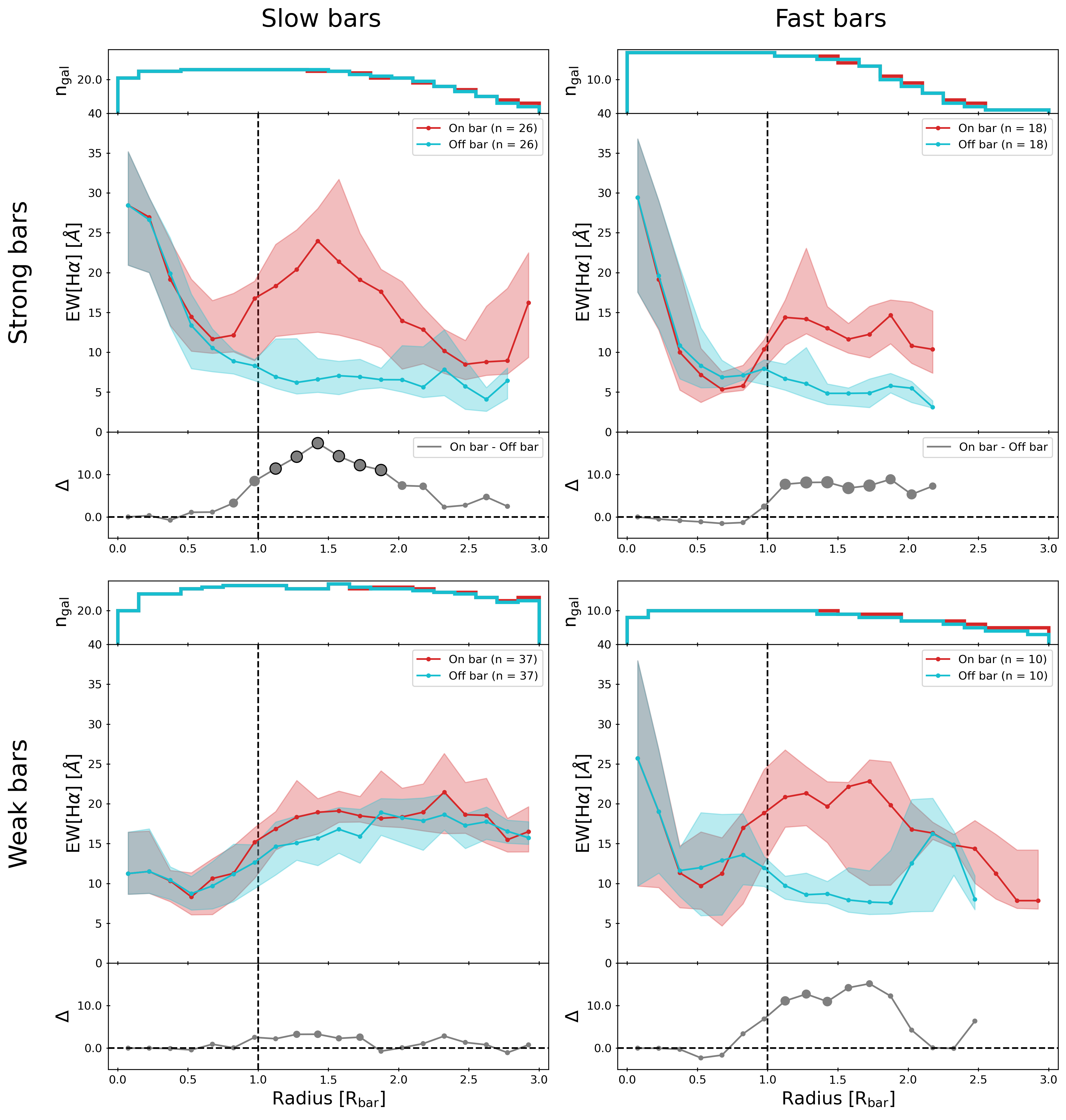}
    \caption{The radial profiles of EW[H$\alpha$] are shown for strongly barred SF galaxies (top row) and weakly barred SF galaxies (bottom row) for slow bars (left column) and fast bars (right column). A radial profile was calculated for every target with the aperture positioned along the bar (red) and perpendicular to the bar (light blue). The radial profiles are shown normalised against the bar radius. The sample size, n, for every sample is indicated in the legend. The solid line is the median in every bin, while the shaded regions bound the 33$^{\rm rd}$ and 66$^{\rm th}$ percentiles. The dashed line at $R_{\textrm{bar}}$ = 1 denotes the end of the bar. The width of the bins is \todo{0.15 $R_{\textrm{bar}}$}. The top part of each panel shows the number of galaxies that have spaxels in each radial bin. The bottom part of each panel shows the difference between each radial profile in every bin, where the size of the point represents the significance of the difference after comparing the two populations with an Anderson-Darling test. The smallest sizes represent a significant difference of less than 1$\sigma$, while the largest sizes represent $>$3$\sigma$ and are additionally outlined in black. This figure shows that a slow strong bar, will increase star formation along the PA of the bar, but not in the entire galaxy.}
    \label{fig:r_profile_ewha_paraperp}
\end{figure*}

This hypothesis can be tested by constructing radial profiles of EW[H$\alpha$] with apertures parallel and perpendicular to the bar and comparing the differences. This is done for all SF galaxies in Figure \ref{fig:r_profile_ewha_paraperp}. The barred galaxies are divided into four subsamples: galaxies with slow strong bars, fast strong bars, slow weak bars and fast weak bars. The biggest difference between the parallel and perpendicular apertures was observed among slow strong bars, where the parallel aperture has significantly ($>$3$\sigma$) higher values for EW[H$\alpha$] between $\sim$1.2 - 1.8 $R_{\rm bar}$ compared to the perpendicular aperture. No significant difference ($>$3$\sigma$) between the parallel and perpendicular apertures is observed for fast strong bars, fast weak or slow weak bars. 

This confirms that \textbf{a slow bar will increase star formation along the position angle of the bar, but not in the entire galaxy.} This was only observed for slow strong bars, which suggests that a slow bar will have more of an effect on its host if it is also a strong bar. This result is further discussed in Section \ref{sec:disc_fbsb_localeffect}.

\subsubsection{Differences in velocity at bar-end}
\label{sec:result_fbsb_velocity}

\begin{figure*}
  \centering
    \includegraphics[width=\textwidth]{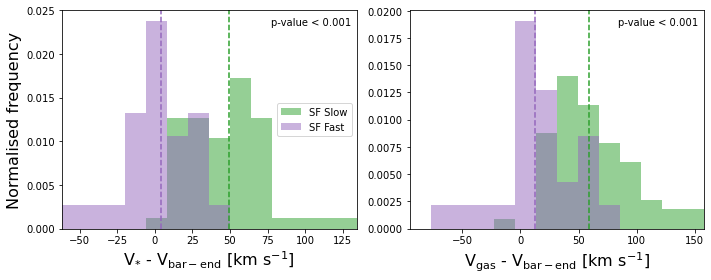}
    \caption{The difference between the velocity of the bar-end region ($V_{\rm bar-end}$) and the velocity of the stars ($V_{*}$) at $R_{\rm bar}$ is shown in the left panel, while the difference between $V_{\rm bar-end}$ and the velocity of the gas ($V_{\rm gas}$) at $R_{\rm bar}$ is shown in the right panel for all slow (green) and fast (purple) SF galaxies. The dashed vertical lines represent the median value of each distribution. The p-value of a two-sample Anderson-Darling test is shown inside each subplot, with the null hypothesis being that the two samples are drawn from the same population. It is clear that the velocity difference is significantly bigger for slow bars than for fast bars.}
    \label{fig:velocity_difference}
\end{figure*}

The results presented in the last few sections suggest that slow bars induce more star formation along the bar compared to fast bars in SF galaxies. However, the underlying physical cause of this increased star formation remains unclear. As mentioned above, the classification of bars into fast and slow depends on $\mathcal{R}$, which is based on the kinematics of the bar and galaxy. Fast bars typically end near the corotation radius, whereas slow bars have bar radii shorter than the corotation radius, given that $\mathcal{R} = R_{\rm CR} / R_{\rm bar}$. The corotation radius is the radius at which the stars in the disc rotate with a speed equal to the rotation of the bar. This means that the bar-end region of fast bars typically rotates with a velocity similar to stars in the disc, whereas the bar-ends of slow bars should rotate much slower. This is illustrated in the left panel of Figure \ref{fig:velocity_difference}, where the difference in the velocity of the stars ($V_{*}$) at $R_{\rm bar}$, based on the stellar rotation curve, and the velocity of the bar-end ($V_{\rm bar-end}$) is measured for all SF galaxies. Bars with $\mathcal{R} > 1$ will have $V_{*} > V_{\rm bar-end}$, while bars with $\mathcal{R} < 1$ (i.e. ultrafast bars) will have $V_{*} < V_{\rm bar-end}$. For bars with $\mathcal{R} = 1$, the velocities will be equal to each other, as $R_{\rm CR} = R_{\rm bar}$ for these bars. The difference in velocity is clearly bigger for slow bars in SF galaxies compared to fast bars in SF galaxies, as indicated by an Anderson-Darling test (p-value $<$ 0.001; $>$3.3$\sigma_{\rm AD}$). The median difference in velocity is \todo{4.2} km s$^{-1}$ for fast bars and \todo{49.6} km s$^{-1}$ for slow bars. 

Note that these differences in velocity are a direct consequence of the definitions of fast and slow bars based on $\mathcal{R}$, so this is what we expect to find. Nevertheless, quantifying this difference highlights how significant it is. Additionally, $\mathcal{R}$ was calculated based on the stellar kinematics of the galaxy. However, we also see a difference between the velocity of the gas ($V_{\rm gas}$) at $R_{\rm bar}$, based on the gas rotation curve, and the velocity of the bar-end ($V_{\rm bar-end}$), which is shown in the right panel of Figure \ref{fig:velocity_difference}. The gas rotation curves are constructed similarly to the stellar rotation curves (see Section \ref{sec:tw_method}). However, instead of using the stellar velocity maps from MaNGA, the gas rotation curve is derived from a Gaussian fit to the H$\alpha$ emission line. The difference between $V_{\rm gas}$ and $V_{\rm bar-end}$ is significantly higher for slow bars in SF galaxies (the median value is \todo{59.0} km s$^{-1}$) than for fast bars (\todo{12.7} km s$^{-1}$) in SF galaxies. 

These results confirm that \textbf{slow bars have a large difference in velocity compared to the stars and gas in the disc in the bar-end region. This is not observed for fast bars}. These differences in velocity at the bar-end between slow and fast bars suggest that slow bars come into contact and interact with a more significant amount gas than fast bars, which could explain the observed differences in local star formation. The implications of this are discussed in more detail in Section \ref{sec:disc_fbsb_velocity}.

\section{Discussion}
\label{sec:discussion}


\subsection{Strong and weak bars}
\label{sec:disc_sbwb}

\subsubsection{Star formation in different regions of strong and weak bars}
\label{sec:disc_sbwb_barregions}

\citet{geron_2021} found that strong bars in SF galaxies increase the central star formation of their hosts, while weak bars do not. However, the impact of strong bars in other regions, and whether weak bars behave similarly in these other regions, remained unclear. This was studied in Figures \ref{fig:r_profile_ewha} and \ref{fig:r_profile_dn4000} using EW[H$\alpha$] and D$_{\rm n}$4000, respectively. These figures show that strongly barred SF galaxies have higher star formation in the centre of their galaxy, compared to weakly barred and unbarred SF galaxies, in agreement with the results from \citet{geron_2021}. There are many other examples in the literature of barred galaxies increasing the SFR in their host \citep{alonso_herrero_2001, hunt_2008, ellison_2011,coelho_2011, hirota_2014,janowiecki_2020,magana_serrano_2020,lin_2020}. The increase in central star formation in strongly barred galaxies can be attributed to the gas inflow from the outskirts to the centre induced by the bar \citep{athanassoula_1992,athanassoula_2013,villa-vargas_2010}, where the gas is available to increase star formation. However, this is only found here for strongly barred SF galaxies and is not observed in weakly barred SF galaxies or all quenching galaxies.


Figures \ref{fig:r_profile_ewha} and \ref{fig:r_profile_dn4000} also show that, on average, the arms of a strong bar on suppress star formation, as suggested by low values for EW[H$\alpha$] and high values for D$_{\rm n}$4000 around $R \approx 0.5 R_{\textrm{bar}}$. This is consistent with other studies that have found evidence for lower SFR and star formation efficiency (SFE) in the arms of the bar \citep{reynaud_1998,sheth_2000,watanabe_2011,khoperskov_2018}. This is typically attributed to high velocity dispersion or shear caused by the bar potential \citep{athanassoula_1992,reynaud_1998,sheth_2000,zurita_2004, haywood_2016,khoperskov_2018} or by fast cloud-cloud collisions \citep{fujimoto_2014b, maeda_2018, fujimoto_2020,maeda_2021}. Regardless of the specific physical process, these findings imply substantial gas flows along the arms of a strong bar. These results are also consistent with the `star formation desert' described by \citet{james_2018} as well as the gas-depleted regions observed in barred galaxies in \citet{george_2019} and \citet{newnham_2020}. The EW[H$\alpha$] and D$_{\rm n}$4000 profiles of weakly barred galaxies are very similar to the profiles of unbarred galaxies in the arms of the bar of SF galaxies. This implies that the suppression of star formation and evidence for gas flows are not found for weak bars in SF galaxies and that weak bars are not as efficient in transporting gas to the centre of the galaxy. This is also in agreement with the previous result that lower rates of star formation are found in the centre of weakly barred galaxies. Interestingly, the EW[H$\alpha$] profile of weakly barred quenching galaxies also drops at $R \approx 1 R_{\textrm{bar}}$, although not as abruptly as in strongly barred quenching galaxies. This suggests that the arms of a weak bar in a quenching galaxy can still suppress some star formation, possibly due to increased velocity dispersion in the barred region of galaxies with lower gas concentrations.


There is a second peak of EW[H$\alpha$] found just beyond the bar-end ($R \approx 1.2-1.5 R_{\textrm{bar}}$) in strongly barred SF galaxies, shown in Figure \ref{fig:r_profile_ewha}. This was studied in more detail in Figure \ref{fig:r_profile_barend_sbwb}, where we found that the difference in EW[H$\alpha$] and D$_{\rm n}$4000 between slits placed parallel and perpendicular to the bar around the bar-end region was larger for strongly barred SF galaxies compared to weakly barred SF galaxies, which illustrates the importance of the bar-end region for strong bar. This is in agreement with other studies that have found increased star formation in the bar-end region \citep{reynaud_1998, verley_2007, emsellem_2015, diaz_garcia_2020}. For example, \citet{maeda_2020b} found higher SFEs in the bar-end region than in the arms of the bar. \citet{fraser_mckelvie_2020} also found increased H$\alpha$ in the bar-ends of 18\% of barred galaxies, most of which have high stellar masses ($M_{*} > 10^{10} M_{\odot}$). They also found increased H$\alpha$ in a ring around the bar, including at the bar-end in an additional 21\% of their sample. This brings the total to 38\% of galaxies in \citet{fraser_mckelvie_2020} with elevated H$\alpha$ emission in the bar-end region. However, we only observe increased H$\alpha$ emission specifically in SF galaxies, while \citet{fraser_mckelvie_2020} did not differentiate between SF and quenching galaxies. This implies that the fraction of galaxies with increased H$\alpha$ emission in the bar-end region in \citet{fraser_mckelvie_2020} would likely be even higher if only SF galaxies were considered. This beyond the bar-end region might correspond to the ``bar shoulders'' found in the simulations of \citet{anderson_2022} and \citet{beraldo_e_silva_2023}. They show that these shoulders form due to looped x$_{1}$ orbits and appear in growing bars. \citet{erwin_2023} also found these shoulders in the surface brightness profiles of nearby barred galaxies. Interestingly, they noted that the shoulders were preferentially found in stronger bars, though they mention that this trend could mostly be explained by accounting for stellar mass. This is consistent with our results, as the increase in EW[H$\alpha$] beyond the bar-end is highest for the galaxies with higher mass (see Figure \ref{fig:r_profile_mass}). These increases in star formation at the bar-end can be attributed to an increased probability of cloud-cloud collisions due to orbital crowding and high gas density, and low amounts of shear in the bar-end region \citep{renaud_2015, emsellem_2015,fraser_mckelvie_2020}. However, an increase in star formation in the bar-end is more pronounced in this work for strongly barred SF galaxies, compared to weakly barred SF galaxies. This suggests that a weak bar less capable of inducing or supporting the effects described above.


The profiles in Figures \ref{fig:r_profile_ewha} and \ref{fig:r_profile_dn4000} show that strongly barred SF galaxies have significantly lower EW[H$\alpha$] and higher D$_{\rm n}$4000, implying lower star formation, in the outskirts of the galaxy, compared to weakly barred and unbarred SF galaxies. This is consistent with the hypothesis that strong bars induce gas inflow from the outskirts to the centre of the galaxy, where it is available to increase star formation \citep{athanassoula_1992,athanassoula_2013}. However, this result is more prominent in the radial profiles expressed in absolute units and less in the ones normalised to the length of the bar. This can be explained by the normalisation that is used: a distance of $2 R_{\textrm{bar}}$ extends much farther out into in the disc of the galaxy for a strong bar than for a weak bar, as strong bars are longer than weak bars.

These results show that strong bars have more star formation in their centre and beyond the bar-end, while suppressing star formation in the arms of the bar. These findings are consistent with much of the literature. This suggests heavy gas flows along the arms of a strong bar to the centre and confirms that strong bars have a significant influence on their host. Additionally, these results show that strong bars can facilitate the quenching process, which is in agreement with the results from \cite{geron_2021}. However, these observations are not found for weakly barred galaxies, which suggest that they do not affect their host in a significant way. This highlights the importance of bar strength on galaxy evolution and quenching.

\subsubsection{Strong bars are long-lived}
\label{sec:disc_sbwb_lifetime}

The shape of the D$_{\rm n}$4000 profiles of strongly barred galaxies are very different to those of weakly and unbarred galaxies, as shown in Figures \ref{fig:r_profile_dn4000} and \ref{fig:r_profile_barend_sbwb}. This implies that strong bars are long-lived structures, as they have been able to able to influence the average age of the stellar populations. This is particularly evident in the arms of the bar. The median value of D$_{\rm n}$4000 for strongly barred SF galaxies in the arms of the bar is \todo{$\sim$1.4}. This suggests that the mean age of the stellar population in the arms of the bar is $\sim$1 Gyr \citep{kauffmann_2003,paulino_afonso_2020}, which is consistent with the results presented in \citet{geron_2023} as well as other studies that show that bars are robust and have a long lifetime \citep{jogee_2004,shen_2004,debattista_2006,kraljic_2012,athanassoula_2013}.

However, this was not observed for weakly barred galaxies, whose D$_{\rm n}$4000 profiles closely resemble those of unbarred galaxies. This could mean that weak bars are transient and short-lived structures that do not have the time to affect the stellar populations of their host. The simulations of \citet{bournaud_2002} and \citet{bournaud_2005} found that bars are short-lived and significantly weaken before they are destroyed. The results presented in this paper suggest that this may be applicable to weak bars, but not to strong bars. Alternatively, weak bars might be simply recently formed structures or they lack the ability influence the stellar population of their host in a significant way, even if they were long-lived.

\subsubsection{Implications for bar continuum}
\label{sec:disc_sbwb_barcontinuum}

The bar continuum was introduced in \citet{geron_2021}, which showed that weak and strong bars are part of a continuum of bar types and that differences between weak and strong bars disappeared when correcting for bar length. The bar continuum suggests that the radial profile of a weak bar should gradually change to that of a strong bar when considering bars of increasing bar strength along this continuum. For example, as the  bar becomes stronger, the increase of EW[H$\alpha$] in the centre and the suppression of EW[H$\alpha$] in the arms of the bar should become increasingly more pronounced. This suggests that there are many intermediate radial profiles between the typical weak and strong profiles. This is indeed what is found (see Figures \ref{fig:r_profile_ewha} and \ref{fig:r_profile_dn4000} in particular), as the shaded regions in the radial profiles indicate that there is substantial variability and scatter in these profiles. An alternative possibility is that the probability of observing these features (i.e. the peak of EW[H$\alpha$] in the centre) increases with bar strength. Either scenario can be tested in future work by characterising the diversity of the radial profiles of galaxies of different bar types in more detail.


\subsection{Fast and slow bars}
\label{sec:disc_fbsb}

\subsubsection{Star formation in different regions of fast and slow bars}
\label{sec:disc_fbsb_barregions}

We investigated the impact of bar kinematics on their host galaxies in terms of star formation in Section \ref{sec:r_profile_fb_sb} using EW[H$\alpha$] and D$_{\rm n}$4000. Figure \ref{fig:r_profile_ewha_fbsb} shows that slow bars in SF galaxies have significantly higher values of EW[H$\alpha$] in the intermediate radius range of the galaxy ($\sim$2-4 kpc) compared to fast bars in SF galaxies. 

This is consistent with the results found in Figure \ref{fig:r_profile_dn4000_fbsb}, where significantly lower values for D$_{\rm n}$4000 were found in the intermediate radius range of SF galaxies with slow bars, compared to fast bars in SF galaxies. This suggests the presence of younger stellar populations due to more recent star formation and implies that slow bars in SF galaxies have significantly more star formation compared to fast bars along the PA of the bar.

This effect is only observed for SF galaxies and not for quenching galaxies. The radial profiles for fast and slow bars in quenching galaxies have similar trends: EW[H$\alpha$] increases with radius, while D$_{\rm n}$4000 decreases monotonically with radius. A similar result was found in \citet{geron_2021}, who showed that strong bars increase the central star formation for SF galaxies, but not for quenching galaxies. This is presumably because the underlying physical processes that cause the observed differences in star formation are related to the gas distribution of the galaxy. If a feature (e.g. a slow bar) influences the gas distribution in a galaxy, and SF galaxies have more gas than quenching galaxies \citep{baldry_2006,dekel_2006}, it is expected that this feature affects SF galaxies more than quenching galaxies.

\subsubsection{Effect of bar strength}
\label{sec:disc_fbsb_barstrength}

The impact of bar strength on the results presented above was studied in Figure \ref{fig:r_profile_fbsb_sbwb_sf} in SF galaxies using EW[H$\alpha$] and D$_{\rm n}$4000. The median value for EW[H$\alpha$] was always higher in slow strong bars than in fast strong bars. The increase of EW[H$\alpha$] beyond the bar-end region, which is characteristic for strong bars, was much more pronounced in slow strong bars compared to fast strong bars, implying that the latter have more star formation there than the former. Additionally, the median EW[H$\alpha$] in the arms of a slow strong bar was also higher than in the arms of a fast strong bar.

The median value for D$_{\rm n}$4000 was also consistently lower along slow strong bars compared to fast strong bars. The value for D$_{\rm n}$4000 was especially high in the arms of fast strong bars, suggesting that the arms of fast strong bars are more efficient in suppressing star formation than the arms of slow strong bars. However, an Anderson-Darling test reveals that these profiles are not significantly different. The profiles for fast weak and slow weak bars are very similar to each other and are not significantly different.

The biggest limiting factor of this analysis is the sample size. These results hint that a slow bar will increase star formation more if it is also a strong bar. In contrast, the profiles of fast weak and slow weak bars are difficult to distinguish from each other. However, a bigger sample size is needed to verify whether these effects are statistically significant. The sample size is relatively modest due to the need for very robust data to perform the TW method. Although this is the largest sample the TW method has been applied to so far, it only contains \todo{210} galaxies with reliable measurements of $\mathcal{R}$ (see \citet{geron_2023} for more details). Subsequently dividing the sample into different subsamples (SF or quenching, weak or strong, fast or slow) further reduces the sample size in the final comparisons. Increasing the total sample size, by including data from past and future IFU surveys, would greatly help to clarify the impact that bar strength has on the profiles of fast and slow bars.

\subsubsection{Local, but not global effect}
\label{sec:disc_fbsb_localeffect}

Even though we have shown that slow bars in SF galaxies have higher EW[H$\alpha$] along the PA of the bar, Figure \ref{fig:sfr_fbsb} shows that the global SFR is not significantly different between SF galaxies with slow or fast bars. This apparent contradiction is explained in Figure \ref{fig:r_profile_ewha_paraperp}, where we constructed radial profiles with apertures placed parallel and perpendicular to the PA of the bar. The EW[H$\alpha$] between the parallel and perpendicular profiles was only significantly different for slow strong bars in SF galaxies, particularly in the region beyond the bar-end (1.2-1.5 $R_{\rm bar}$). This is also the region where the second peak of EW[H$\alpha$] was typically found in strongly barred SF galaxies in Figure \ref{fig:r_profile_ewha}, implying that this peak is most prevalent among strong bars in SF galaxies that are also slow. Although not significantly different, the median value of EW[H$\alpha$] for the profile with the parallel aperture is higher than the profile with the perpendicular aperture in the region beyond the bar-end of fast strong and fast weak bars as well. The profiles for the parallel and perpendicular apertures for weak strong bars in SF galaxies were also not significantly different. This implies that the distribution of star formation along the PA of the bar is shaped by whether the bar is fast or slow, despite both fast and slow bars having similar effects on their host in terms of global SFR. This suggests that star formation is more concentrated along the bar for slow bars compared to fast bars. However, as the global SFR remains the same, this also suggests that a slow bar does not change the overall rate of gas consumption of its host. Instead, a slow bar influences its host by changing where gas consumption and star formation take place. As a side note, these results also highlight the importance of using IFUs to study galaxy evolution. The crucial kinematic distinction between fast and slow bars could not have been found using only global properties and photometry.

\subsubsection{Effect of velocity}
\label{sec:disc_fbsb_velocity}

The results from the previous section suggest that slow bars have more star formation along their bars. But we have not yet looked into a possible physical explanation for this observation. A bar is classified as slow or fast based on its kinematics. Slow bars are shorter than their corotation radius, whereas fast bars end near the corotation radius. As the corotation radius is the radius where the stars in the disc rotate with the same speed as the bar, this implies that the bar-end of fast bars should move with a similar velocity as the stars in the disc. Conversely, the stars in the disc should rotate faster than the bar-end region of slow bars. This is confirmed in the left panel of Figure \ref{fig:velocity_difference}, while the right panel shows that this is also the case for the gas in the disc: the difference in velocity between the bar-end and gas in the disc is much higher for slow bars than for fast bars. 

\begin{figure*}
  \centering
    \includegraphics[width=\textwidth]{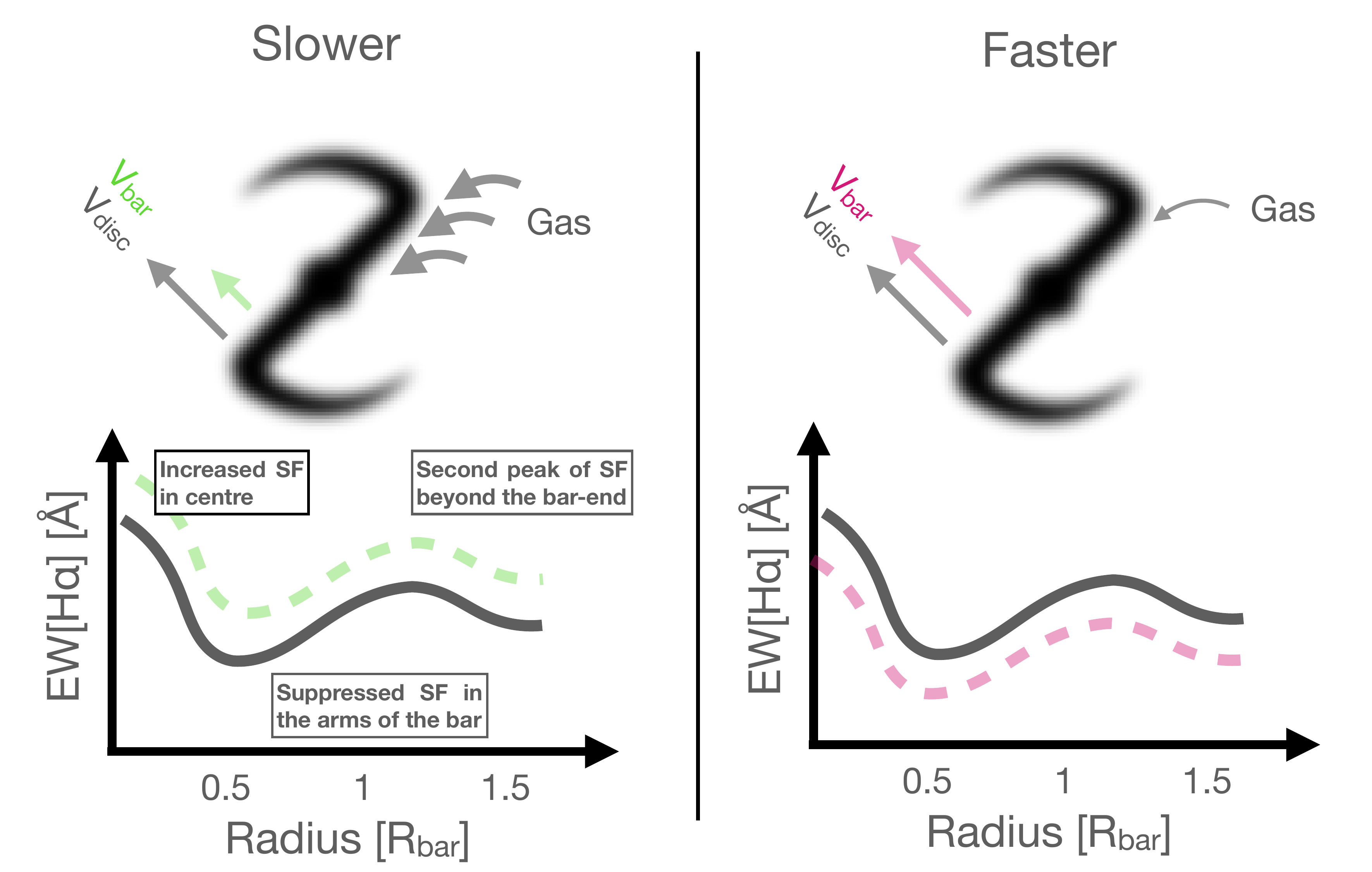}
    \caption{This toy model illustrates a possible physical interpretation of the results presented in this work. The left panel shows how the radial profile of a strongly barred galaxy will change if the bar is also a slow bar. The bar-ends of slow bars rotate with a lower velocity than the stars and gas in the disc of the galaxy. Consequently, a slow bar will come into contact with much more of the gas of its host compared to a fast bar. This will cause more gas to be trapped in the slow bar, which will flow along the bar to the centre of the galaxy. This increased concentration of gas along the bar will increase the observed values of EW[H$\alpha$] in the radial profiles of slow bars. Conversely, the right panel shows what happens for fast bars. The bar-ends of a fast bar rotate with a similar speed to the stars and gas in the disc. This will cause less gas to be trapped in the bar and lower values of EW[H$\alpha$] along the bar, compared to slow bars.} 
    \label{fig:cartoon}
\end{figure*}

Figure \ref{fig:cartoon} shows a possible interpretation of how these differences in velocity cause changes to the radial profiles of slow bars (left panel) and fast bars (right panel). Because of the greater difference in velocity between the bar-end and disc, a rotating slow bar will come into contact with much more of the gas of its host, compared to a fast bar. This suggests that slow bars possibly trap and concentrate more gas along the bar than fast bars. This increased concentration of gas will increase the observed values of EW[H$\alpha$] along the bar for slow bars. Conversely, the bar-ends of fast bars rotate with a similar velocity to the stars and gas in the disc and will not concentrate as much gas along the bar. This results in lower observed values of EW[H$\alpha$] along the bar for fast bars. We stress that this is one possible interpretation of the results presented in this paper and that more detailed observations of the effect of bar kinematics on the distribution of gas will help to clarify any remaining questions.

This interpretation suggests that slow bars are possibly more efficient at sweeping up gas and creating gas-depleted regions, such as the ones observed by \citet{gavazzi_2015,james_2018,george_2019, newnham_2020}. Simulations also find such gas-depleted regions. For example, the high-resolution cosmological `zoom-in' simulation of \citet{spinoso_2017} finds a very clear gas-depleted region in a barred galaxy. The bar has a radius of $\sim$1.5 kpc at z = 0.02, at which point the corotation radius is $\sim$3.5 kpc, so that $\mathcal{R}$ $\approx$ 2.3. Therefore, the bar that created the obvious gas-depleted region is a slow bar. Similarly, the self-consistent simulations of Milky Way-sized isolated disc galaxies in \citet{seo_2019} also show clear gas-depleted region in barred galaxies. The bars all start out as fast, but quickly become slow. However, as explained in \citet{geron_2023}, simulations tend to overestimate the observed values for $\mathcal{R}$, which implies that studying the effect that slow and fast bars have on the gas-depleted regions using simulations might not be robust. It would be better to study this observationally instead. This could be done with resolved gas observations of multiple slow and fast bars in galaxies, verifying whether the gas-depleted region is more pronounced in slow bars than in fast bars.





However, as noted in the previous section and shown in Figures \ref{fig:sfr_fbsb} and \ref{fig:r_profile_ewha_paraperp}, slow and fast bars have similar global SFRs, despite the differences observed along the bar. This suggests that the total star formation efficiency of the galaxies remains the same. Even though a slow bar is likely to be more efficient at sweeping up and concentrating gas in the bar region compared to a fast bar, an additional element must be acting to counteract the expected increase in star formation efficiency that a higher gas concentration implies. A likely candidate is the increased velocity dispersion or shear that is commonly associated with strong gas flows in the arms of the bar \citep{athanassoula_1992,reynaud_1998,sheth_2000,zurita_2004, haywood_2016,khoperskov_2018}. As Figure \ref{fig:r_profile_fbsb_sbwb_sf} shows, the arms of the bar are where the most significant differences are observed between fast and slow bars, especially if the bar is strong as well. Thus, while slow bars increase the concentration of gas in the bar, the global SFR does not increase due to the high amounts of shear in the arms of the bar. These opposing effects result in a global SFR in slow bars that is similar to that of fast bars. In other words, because of these effects, the kinematics of the bar do not affect global SFR, but dictate where star formation occurs.

\subsubsection{A `kinematic' bar continuum}
\label{sec:disc_fbsb_kinematiccontinuum}

The idea of fast and slow bars works conceptually well with the continuous nature of bar types, given that  $\mathcal{R}$, which is used to classify bars into slow or fast, is a continuous variable. Futhermore, $\mathcal{R}$ is defined as the ratio of the corotation radius to the bar radius, both of which are known to change over time. This implies that $\mathcal{R}$ can vary over time as well and that bars can become `slower' or `faster'. This is similar to the classification of bars into strong and weak, based on $p_{\rm strong}$. This parameter can also change over time, so that bars can become stronger or weaker. However, the classification of bars into weak or strong, and the bar continuum presented in \citet{geron_2021}, are based on visual morphology. In contrast, the classification of bars as fast and slow is based on the kinematics of the bar. This suggests that a potential second - kinematic - axis should be added to the bar continuum, which would accurately reflect that the kinematics of the bar has an important role in galaxy evolution as well. 

\begin{figure*}
    \centering
    \includegraphics[width=0.7\textwidth]{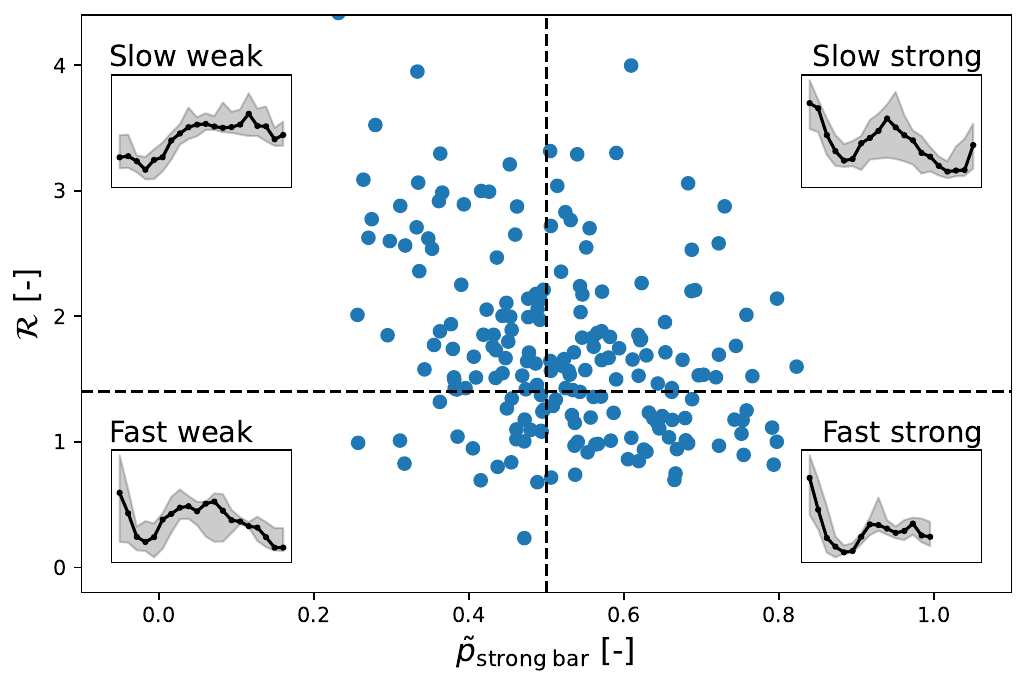}
    \caption{Visualisation of how the targets in the TW sample would lie on a two-dimensional bar continuum, with one axis based on morphological arguments (horizontal axis, $\tilde{p}_{\textrm{strong bar}}$) and one axis based on kinematic arguments (vertical axis, $\mathcal{R}$). The dashed vertical line separates strong and weak bars, while the dashed horizontal line separates (ultra)fast and slow bars. This means that there are four distinct quadrants in this plot for the different bar types: slow weak bars, fast weak bars, fast strong bars and slow strong bars. The characteristic radial profile of EW[H$\alpha$] of SF galaxies in each quadrant (as calculated in Figure 
    \ref{fig:r_profile_fbsb_sbwb_sf}) is shown in the corresponding insets. The Spearman correlation coefficient equals \todo{-0.31}, and its significance is \todo{4.63}$\sigma$.}
    \label{fig:2d_continuum}
\end{figure*}

A visualisation of the distribution of galaxies in the TW sample on such a two-dimensional plane is shown in Figure \ref{fig:2d_continuum}, where $\mathcal{R}$ is plotted against the normalised strong bar vote fraction ($\tilde{p}_{\textrm{strong bar}}$) from GZ DESI. The normalised strong bar vote fraction is defined as follows:
\begin{equation}
    \tilde{p}_{\textrm{strong bar}} = \frac{p_{\textrm{strong bar}}}{p_{\textrm{weak bar}} + p_{\textrm{strong bar}}}\;,
    \label{eq:p_strong_normalised}
\end{equation}
where $p_{\textrm{strong bar}}$ is the strong bar vote fraction from GZ DESI and $p_{\textrm{weak bar}}$ the weak bar fraction fraction from GZ DESI (see Section \ref{sec:galaxy_zoo} for more detailed information about Galaxy Zoo). This is done to eliminate the dependence of the unbarred vote fraction ($p_{\textrm{no bar}}$) and implies that all strong bars have $\tilde{p}_{\textrm{strong bar}} > 0.5$ and all weak bars have $\tilde{p}_{\textrm{strong bar}} < 0.5$. 

This plane has four separate quadrants: one for galaxies with slow weak bars, fast weak bars, fast strong bars and slow strong bars. The characteristic radial profile of EW[H$\alpha$] of SF galaxies in each quadrant (as calculated in Figure 
\ref{fig:r_profile_fbsb_sbwb_sf}) is shown in the corresponding insets. It is clear from this figure that there are barred galaxies in every quadrant. This means that a classification based on bar strength does not preclude any classification based on kinematics. In order words, a strong bar can still be either fast or slow. This also suggests that both the kinematic and photometric information are needed in order to fully characterise a barred galaxy.

The results in this paper suggest that barred galaxies on the right side of the plot (strong bars, $\tilde{p}_{\textrm{strong bar}} > 0.5$) will increase the star formation in the centre of the bar and beyond the bar-end, while suppressing star formation along the arms of the bar. This is in contrast to barred galaxies on the left side of the plot (weak bars, $\tilde{p}_{\textrm{strong bar}} < 0.5$), which do not seem to significantly influence star formation in their hosts. Additionally, barred galaxies in the upper half of the plot (slow bars, $\mathcal{R} > 1.4$) will concentrate their star formation more along their barred regions and sweep up gas more efficiently, compared to barred galaxies in the bottom half of this plot (fast bars, $\mathcal{R} < 1.4$). Therefore, bars in the upper-right quadrant of Figure \ref{fig:2d_continuum}, i.e. slow strong bars, will affect their hosts the most.


\section{Conclusions}
\label{sec:r_profile_conclusions}

In this paper, we have created radial profiles of EW[H$\alpha$] and D$_{\rm n}$4000 for all the strongly, weakly and unbarred galaxies in the GZ DESI-MaNGA sample in Section \ref{sec:r_profile_sb_wb} and for all the fast and slow bars in the TW sample in Section \ref{sec:r_profile_fb_sb} using IFU data from MaNGA. These profiles were used to study how bar strength (strong and weak bars) and bar kinematics (fast and slow bars) affect star formation in different regions of the galaxy.

Previous work showed that strong bars have increased central star formation in SF galaxies, compared to weak bars \citep{geron_2021}. However, they did not study what happened in different regions of the bar, such as the arms of the bar and the bar-end regions. The kinematics of barred galaxies was studied in \citet{geron_2023}, who identified slow ($\mathcal{R} > 1.4$), fast ($1.0 < \mathcal{R} < 1.4$) and ultrafast bars ($\mathcal{R} < 1.0$). However, it was still unclear whether fast and slow bars affect their host differently in terms of star formation and galaxy quenching. These issues were addressed in this paper.

\begin{enumerate}
    \item \textit{\textbf{Strong and weak bars:}}
    
    \begin{itemize}
        \item We found more star formation in the centre and beyond the bar-end region of strongly barred SF galaxies, compared to weakly barred and unbarred SF galaxies. In contrast, there is less star formation in the arms of the bar and the outskirts of the galaxy in strongly barred SF galaxies. These observations can be explained by heavy gas flows induced by the bar, which flow along the arms of the bar to the centre of the galaxy, where the gas is available for star formation. This suggests that strong bars can significantly affect the evolution of galaxies.
        \item In contrast, the EW[H$\alpha$] and D$_{\rm n}$4000 profiles of weakly barred SF galaxies were largely similar to those of unbarred SF galaxies, suggesting that weak bars do not induce significant gas flows and do not have the ability to significantly affect their host.
        \item These results remained valid in intermediate and high mass galaxies. However, the suppression of star formation in the arms of the bar and the increase of star formation in the bar-end for strongly barred SF galaxies were no longer observed among low mass galaxies, suggesting that stellar mass plays an important role in galaxy quenching. However, the increase in central star formation in strongly barred SF galaxies was still observed in low mass galaxies.
        \item Strong bars are long-lived structures, as they can influence the age of the stellar population, as shown by the D$_{\rm n}$4000 profiles. This was not found for weak bars, whose D$_{\rm n}$4000 profiles are similar to those of unbarred galaxies.
        \item The median radial profiles of weakly and strongly barred galaxies are found to be very different. However, the shaded regions around these median profiles indicate that there is still large amount of variability, which suggests that there are many intermediate profiles. This is consistent with the idea of the bar continuum introduced in \citet{geron_2021}.
    \end{itemize}

    \item \textit{\textbf{Fast and slow bars:}}
    \begin{itemize}
        \item Slow bars have significantly higher star formation along the bar than fast bars in SF galaxies, indicated by the EW[H$\alpha$] and D$_{\rm n}$4000 profiles.
        \item However, we found that the global star formation rate is similar between fast and slow bars, which implies that the total rate of gas consumption in fast and slow bars is still similar. We found that slow bars increase star formation locally along not bar, but not globally. Thus, the kinematics of the bar dictate where star formation occurs.
        \item This increase in star formation along slow bars is most likely caused by larger differences in velocity between the bar-end and the gas in the disc observed in slow bars, compared to fast bars. This implies that slow bars are likely more efficient at sweeping up and concentrating gas in the bar and creating gas-depleted regions.
        \item The work presented in this section suggests that the distinction between fast and slow bars is physically meaningful and that they affect their host in different ways. 
    \end{itemize}

    \item \textit{\textbf{Combining bar strength with bar kinematics:}}
    \begin{itemize}
        \item The results presented here suggest that there is a combined effect between bar strength and bar kinematics: a bar will affect its host the most, in terms of star formation, if it is both strong and slow.
        \item As $\mathcal{R}$ is a continuous variable, bars can become `faster' and `slower'. This is consistent with the continuous nature of bar types from `weakest' to `strongest'. However, the bar continuum presented in \citet{geron_2021} was based on visual morphology, whereas the distinction between fast and slow is based on kinematics. Furthermore, a strong (or weak) bar can be either fast or slow. These results suggest that a possible second axis could be added to the bar continuum, so that it has one morphological axis and one kinematic axis.
    \end{itemize}
    
\end{enumerate}

We have shown that it is important to consider both the kinematics and morphology of bars when studying galaxy evolution, as both separately have different and significant effects on how the bar influences its host. Nevertheless, these intriguing results need to be confirmed by a larger sample size, which will help to improve our understanding of the combined effects of bar kinematics and bar strength.


\section*{Acknowledgements}

The Dunlap Institute is funded through an endowment established by the David Dunlap family and the University of Toronto.

RJS gratefully acknowledges funding from the Royal Astronomical Society.

ILG acknowledges support from an STFC PhD studentship [grant number ST/T506205/1] and from the Faculty of Science and Technology at Lancaster University.

The data in this paper are the result of the efforts of the Galaxy Zoo volunteers, without whom none of this work would be possible.

The DESI Legacy Imaging Surveys consist of three individual and complementary projects: the Dark Energy Camera Legacy Survey (DECaLS), the Beijing-Arizona Sky Survey (BASS), and the Mayall z-band Legacy Survey (MzLS). DECaLS, BASS and MzLS together include data obtained, respectively, at the Blanco telescope, Cerro Tololo Inter-American Observatory, NSF’s NOIRLab; the Bok telescope, Steward Observatory, University of Arizona; and the Mayall telescope, Kitt Peak National Observatory, NOIRLab. NOIRLab is operated by the Association of Universities for Research in Astronomy (AURA) under a cooperative agreement with the National Science Foundation. Pipeline processing and analyses of the data were supported by NOIRLab and the Lawrence Berkeley National Laboratory (LBNL). Legacy Surveys also uses data products from the Near-Earth Object Wide-field Infrared Survey Explorer (NEOWISE), a project of the Jet Propulsion Laboratory/California Institute of Technology, funded by the National Aeronautics and Space Administration. Legacy Surveys was supported by: the Director, Office of Science, Office of High Energy Physics of the U.S. Department of Energy; the National Energy Research Scientific Computing Center, a DOE Office of Science User Facility; the U.S. National Science Foundation, Division of Astronomical Sciences; the National Astronomical Observatories of China, the Chinese Academy of Sciences and the Chinese National Natural Science Foundation. LBNL is managed by the Regents of the University of California under contract to the U.S. Department of Energy. The complete acknowledgments can be found at https://www.legacysurvey.org/acknowledgment/.

Funding for the Sloan Digital Sky Survey IV has been provided by the Alfred P. Sloan Foundation, the U.S. Department of Energy Office of Science, and the Participating Institutions. SDSS acknowledges support and resources from the Center for High-Performance Computing at the University of Utah. The SDSS web site is www.sdss4.org.

SDSS is managed by the Astrophysical Research Consortium for the Participating Institutions of the SDSS Collaboration including the Brazilian Participation Group, the Carnegie Institution for Science, Carnegie Mellon University, Center for Astrophysics | Harvard \& Smithsonian (CfA), the Chilean Participation Group, the French Participation Group, Instituto de Astrof\'isica de Canarias, The Johns Hopkins University, Kavli Institute for the Physics and Mathematics of the Universe (IPMU) / University of Tokyo, the Korean Participation Group, Lawrence Berkeley National Laboratory, Leibniz Institut f\"ur Astrophysik Potsdam (AIP), Max-Planck-Institut f\"ur Astronomie (MPIA Heidelberg), Max-Planck-Institut f\"ur Astrophysik (MPA Garching), Max-Planck-Institut f\"ur Extraterrestrische Physik (MPE), National Astronomical Observatories of China, New Mexico State University, New York University, University of Notre Dame, Observat\'ario Nacional / MCTI, The Ohio State University, Pennsylvania State University, Shanghai Astronomical Observatory, United Kingdom Participation Group, Universidad Nacional Aut\'onoma de M\'exico, University of Arizona, University of Colorado Boulder, University of Oxford, University of Portsmouth, University of Utah, University of Virginia, University of Washington, University of Wisconsin, Vanderbilt University, and Yale University.

This project makes use of the MaNGA-Pipe3D dataproducts. We thank the IA-UNAM MaNGA team for creating this catalogue, and the Conacyt Project CB-285080 for supporting them. 

\bibliography{bibtex}{}
\bibliographystyle{aasjournal}



\end{document}